\documentclass[a4paper]{article}
\usepackage{epsfig}
\usepackage{rotating}
\DeclareGraphicsExtensions{.ps,.eps,.pdf}
\usepackage{amsmath,amssymb,amsthm}
\usepackage{dsfont}
\usepackage{hyperref}
\usepackage{gastex}
\usepackage{triotex}

\newcommand{\tstart}{t_{\mathrm{start}}}

\newcommand{\tait}[1]{\mathrm{#1}}
\newcommand{\idle}{\tait{idle}}
\newcommand{\try}{\tait{try}}
\newcommand{\s}{\tait{s_1}}
\newcommand{\ok}{\tait{ok_1}}
\newcommand{\ko}{\tait{ko_1}}

\newcommand{\okk}{\tait{ok_2}}
\newcommand{\koo}{\tait{ko_2}}
\newcommand{\toutt}{\tait{tout_2}}

\newcommand{\frf}[1]{(\ref{#1})}
\newcommand{\fsrf}[2]{(\ref{#1}--\ref{#2})}

\newcommand{\pp}{\mathsf{p}}

\newcommand{\system}{\phi^{\mathsf{sys}}}
\newcommand{\prop}{\phi^{\mathsf{prop}}}

\newcommand{\mtlplus}{MTL$^+$}
\newcommand{\mtlstar}{MTL$^*$}

\newcommand{\zot}{$\integers$ot}
\newcommand{\tazot}{\textsf{TA}$\integers$ot}

\newcommand{\overap}[1]{\mathrm{O}_{\delta}\left({#1}\right)}

\newcommand{\underap}[1]{\Omega_{\delta}\left({#1}\right)}

\newcommand{\logictrue}{\top}
\newcommand{\logicfalse}{\bot}

\newcommand{\Ical}{\mathcal{I}}
\newcommand{\Jcal}{\mathcal{J}}
\newcommand{\Pcal}{\mathcal{P}}
\newcommand{\Dcal}{\mathcal{D}}
\newcommand{\Bchi}{\mathcal{B}_{\chi}}

\newcommand{\modelstime}[1]{\models_{#1}}

\newcommand{\mdt}{\modelstime{\timedomain}}
\newcommand{\behav}{\mathcal{B}}

\newcommand{\iit}{\mathsf{it}}
\newcommand{\IIT}{\Ical}
\newcommand{\st}{\mathsf{st}}
\newcommand{\inpt}{\mathsf{in}}
\newcommand{\rest}[1]{\mathsf{rs}_{#1}}

\newcommand{\becomesMTL}[1]{\MTLoperator{\triangle}{}{}{#1}}
\newcommand{\becomesOMTL}[1]{\MTLoperator{\blacktriangle}{}{}{#1}}
\newcommand{\becomesLMTL}[1]{\becomesOMTL{#1}}
\newcommand{\XiL}{\overrightarrow{\Xi}}

\theoremstyle{plain}
\newtheorem{trio-axiom}{Axiom}
\newtheorem{trio-theorem}[trio-axiom]{Theorem}
\newtheorem{trio-assumption}[trio-axiom]{Assumption}

\newtheorem{theorem}{Theorem}
\newtheorem{proposition}[theorem]{Proposition}

\theoremstyle{definition}

\title{Practical Automated Partial Verification \\ of Multi-Paradigm Real-Time Models}

\author{Carlo A. Furia, Matteo Pradella, and Matteo Rossi}

\date{April 2008}

\begin{document}

\maketitle

\vspace{3cm}
\begin{abstract}
This article introduces a fully automated verification technique that permits to analyze real-time systems described using a continuous notion of time and a mixture of operational (i.e., automata-based) and descriptive (i.e., logic-based) formalisms.
The technique relies on the reduction, under reasonable assumptions, of the continuous-time verification problem to its discrete-time counterpart.
This reconciles in a viable and effective way the dense/discrete and operational/descriptive di\-chot\-o\-mies that are often encountered in practice when it comes to specifying and analyzing complex critical systems.
The article investigates the applicability of the technique through a significant example centered on a communication protocol. More precisely, concurrent runs of the protocol are formalized by parallel instances of a Timed Automaton, while the synchronization rules between these instances are specified through Metric Temporal Logic formulas, thus creating a multi-paradigm model.
Verification tests run on this model using a bounded validity checker implementing the technique show consistent results and interesting performances.
\end{abstract}

\newpage

\tableofcontents

\newpage

\section{Introduction}
There is a tension between the standpoints of modeling and of verification when it comes to choosing a formal notation.
The ideal modeling language would be very expressive, thus capturing sophisticated features of systems in a natural and straightforward manner; in particular, for concurrent and real-time systems, a dense time model is the intuitive choice to model true asynchrony seamlessly.
On the other hand, expressiveness is often traded off against complexity (and decidability), hence the desire for a feasible and fully automated verification process pulls in the opposite direction of more primitive, and less expressive, models of time and systems.
Discrete time, for instance, is usually more amenable to automated verification, and quite mature techniques and tools can be deployed to verify systems modeled under this assumption.

Another, orthogonal, concern of the real-time modeler is the choice between operational and descriptive modeling languages.
Typical examples of operational notations are Timed Automata (TA) and Timed Petri Nets, while temporal logics are popular instances of descriptive notations.
Operational and descriptive notations have complementary strengths and weaknesses.
For instance, temporal logics are very effective for describing partial models or requirements about the past (through the natural use of past operators); automata-based notations, on the other hand, model systems through the notions of state and transition, and are typically easy to simulate and visualize.
Hence, from a modeling viewpoint, the possibility of integrating multiple modeling paradigms in formalizing a system would be highly desirable.

This paper introduces a verification technique that, under suitable assumptions, reconciles the dense/discrete and operational/descriptive dichotomies in an effective way.
More precisely: (1) it permits to analyze continuous-time models using fully automated, discrete-time verification techniques; and (2) it allows users to mix operational (TA) and descriptive (metric temporal logic, MTL) components in the system specification.
The technique is partial in two respects: it can fail to provide conclusive answers, and only dense-time behaviors with bounded variability are verified.
It involves an automated translation of the operational part into temporal logic notation, based on an MTL axiomatization discussed in this paper.
The resulting MTL model, describing both the system and the properties to be verified, is then discretized according to the techniques introduced in \cite{FPR08-FM08}.
The discrete-time approximation can be analyzed through conventional tools; we provide an implementation based on the \zot{} bounded satisfiability checker \cite{zot}.

We experimented with a significant example based on the description of a communication protocol by means of a timed automaton.
Concurrent runs of the protocol are formalized by parallel instances of the same automaton; additionally, the simple synchronization rules between these instances is naturally formalized by means of additional MTL formulas, hence building a mixed model.
Verification tests run on these models showed consistent results, and acceptable performances.

An interesting auxiliary contribution of the discretizable axiomatization of TA in MTL is a set of ``rules of thumb'' about how to describe systems based on the notion of state and transition with a logic formalism, in a way which is also amenable to discretization (according to the notion of \cite{FPR08-FM08}).
Section \ref{sec:TA-approx} discusses this issue with great detail.

Finally, let us stress that our approach aims at providing a \emph{practical} approach to the verification of operational (and mixed) models.
Hence, we sacrifice completeness in order to have a lightweight and flexible technique.
Also note that, although in this paper TA are the operational formalism of choice, the same approach could be applied to other operational formalisms, such as Timed Petri Nets.

\paragraph{Structure of the paper.}
The paper is organized as follows.
Section \ref{sec:overview} provides a sketch of the whole technique with as little technical details as possible.
Section \ref{sec:relatedword} briefly summarizes some research related to the content of this paper.
Section \ref{sec:preliminaries} introduces the technical definitions that are needed in the remainder, namely the syntax and semantics of MTL and TA, and the discretization techniques from \cite{FR06,FPR08-FM08} that will be used.
Section \ref{sec:formalizingTAs} shows how to formalize the behavior of TA as a set of dense-time MTL formulas.
Then, Section \ref{sec:TA-approx} re-examines the axioms and suitably modifies them in a way which is most amenable to the application of the discretization technique; the overall result is a set of discrete-time MTL formulas whose satisfiability is linked to the satisfiability of the original dense-time formulas according to the rules of the discretization technique.
Section \ref{sec:impl-example} describes the example of a simple communication protocol and reports on the experiments conducted on it with the SAT-based implementation of the technique.
Finally, Section \ref{sec:conclusions} draws some conclusions.

\subsection{Overview} \label{sec:overview}
The goal of our technique is to provide a means to carry out practical verification technique of real-time systems described using a dense notion of time and a mixture of operational and descriptive notations.
In particular, we assume a model of real time based on the notion of \emph{behavior}, which is basically a continuous-time signal, and we consider a variant of TA as operational formalism and MTL as descriptive formalism.

The most common approaches to similar verification problems involve translating the logic into automata \cite{AFH96}.
In this paper we take the mirror approach of describing TA through MTL formulas.
This choice is mainly justified by the fact that logic formulas are naturally compositional, hence our ultimate goal of formally combining mixed models is facilitated by this choice.
It is well-known that MTL is undecidable over dense time \cite{AH93}; this hurdle is however practically mitigated by employing the \emph{discretization} technique for MTL introduced --- and demonstrated to be practically appealing --- in \cite{FPR08-FM08}.
Note that the undecidability of dense-time MTL entails that the reduction technique must be incomplete, i.e., there are cases in which we are unable to have a conclusive outcome to the verification problem.
However, as demonstrated in \cite{FPR08-FM08}, and further shown here, the impact of this shortcoming can be rendered small in many practical cases.

We start by providing a dense-time MTL axiomatization of TA.
Notice that, due to a well-known expressiveness gap between temporal logics and automata \cite{HRS98} it is impossible to describe the language accepted by a generic TA as an MTL formula.
What we provide is instead a formal description of \emph{accepting runs} of a TA as an MTL formula; in other words, we model the overall behavior of TA with a set of MTL axioms.
The resulting MTL axioms are discretized according to the rules provided in \cite{FPR08-FM08}.
We show that this yields poor results if done na{\"\i}vely; hence, we carefully revise the axiomatization and put it in a way which is much more amenable to discretization.
The result is a set of discretized MTL axioms describing TA runs.
These axioms can be combined with additional pieces of specification, written in MTL, and with the properties to be verified.
The resulting complete model can then be analyzed by means of automated discrete-time tools; the results of the discrete-time analysis are then used, as defined in \cite{FPR08-FM08}, to finally infer results about the verification of the original dense-time model.
The experimental results are encouraging, both in terms of performances and in terms of ``completeness coverage'' of the method.

In this paper we justify the soundness of the technique, which requires several analyses of the axiomatization and of the discretizations that are produced.
It is important to understand, however, that the resulting technique (and tool) is completely automated, and the user has just to provide the dense-time model of the system (i.e., TA and MTL formulas) and the putative properties to be verified.

\subsection{Related Work} \label{sec:relatedword}
To the best of our knowledge, our approach is rather unique in trying to combine operational and descriptive formalisms over dense time, then trading-off verification completeness against better performance and practical verification results.
On the other hand, each of the ``ingredients'' of our method has been studied in isolation in the literature.
In this section we briefly recall a few of the most important results in this respect.

Dense-time verification of operational models is a very active field, and it has produced a few high-performance tools and methods.
Let us mention, for instance, Uppaal \cite{LPY97}, Kronos \cite{Yov97}, HyTech \cite{HHW97}, and PHAVer \cite{Fre05} for the verification of timed (and hybrid) automata.
Notice that, although tools such as Uppaal allow the usage of a descriptive notation to express the properties to be verified, the temporal logic subset is very simple and of very limited expressive power.
In contrast, we allow basically full MTL to be freely used in both the description of the model and in the formalization of the properties to be verified, at the price of sacrificing completeness of verification.

Metric temporal logic (MTL) verification is also a well-understood research topic.
MTL is however known to be undecidable over dense time domains \cite{AH93}.
A well-known solution to this limitation restricts the syntax of MTL formulas to disallow the expression of exact (i.e., punctual) time distances \cite{AFH96}.
The resulting logic, called MITL, is fully decidable over dense time.
However, the associated decision procedures are rather difficult to implement in practice and, even if recently significant progress has been made in simplifying them \cite{MNP06}, a serviceable implementation is still lacking.

Another stance at working around the undecidability of dense-time MTL builds upon the fact that the same logic is decidable over discrete time.
Hence, a few approaches introduce some notion of discretization, that is partial reduction of the verification problem from dense to discrete time.
The present paper goes in this direction by extending previous work on MTL \cite{FPR08-FM08} to the case of TA.
A different discretization technique, based on the notion of robust satisfiability of MTL specifications, has been introduced in \cite{FP07}.
Other work also deals with notions of robustness in order to guarantee that dense-time TA are implementable with non-ideal architectures \cite{DWDR05}.
Another well-known notion of discretization is the one based on the concept of \emph{digitization} \cite{HMP92}; several authors have applied this quite general notion to the practical verification of descriptive \cite{Oua02,VHG96,CP03,SPC05} or operational \cite{GPV94,KP05,BER94,Bos99,MP95,BMT99,BLN03,OW03,CLT07} formalisms.
See also the related work section of \cite{FPR08-FM08} for more references about discretization techniques.

\section{Preliminaries and Definitions} \label{sec:preliminaries}

\subsection{Behaviors}
Real-time system models describe the temporal behavior of some basic items and propositions, which represent the observable ``facts'' of the system.
More precisely, an item $\iit$ is characterized by a finite domain $\Dcal^{\iit}$ (and we write $\iit: \Dcal^{\iit}$) such that at any instant of time $\iit$ takes one of the values in $\Dcal^{\iit}$.
On the other hand, a proposition $\pp$ is simply a fact which can be true or false at any instant of time.

A \emph{behavior} is a formal model of a \emph{trace} (or \emph{run}) of some real-time system.
Given a time domain $\timedomain$, a finite set $\Pcal$ of atomic propositions, and a finite set of items $\IIT$, a behavior $b$ is a mapping $b: \timedomain \rightarrow \Dcal^{\iit_1}\times \Dcal^{\iit_2} \times \cdots \times \Dcal^{\iit_{|\IIT|}} \times 2^\Pcal$ which associates with every time instant $t \in \timedomain$ the tuple $b(t) = \langle v_1, v_2, \ldots, v_{|\IIT|}, P \rangle$ of item values and propositions that are true at $t$.
$\behav_{\timedomain}$ denotes the set of all behaviors over $\timedomain$, for an implicit fixed set of items and propositions.

$b(t)|_{\iit}$ and $b(t)|_{\Pcal}$ denote the projection of the tuple $b(t)$ over the component corresponding to item $\iit$ and the set of propositions in $2^\Pcal$ respectively.
Also, $t \in \timedomain$ is a \emph{transition point} for behavior $b$ if $t$ is a discontinuity point of the mapping $b$.

Whether $\timedomain$ is a discrete, dense, or continuous set, we call a behavior over $\timedomain$ discrete-, dense-, or continuous-time respectively.
In this paper, we consider the natural numbers $\naturals$ as discrete-time domain and the nonnegative real numbers $\reals_{\geq 0}$ as continuous-time (and dense-) time domain.

\paragraph{Non-Zeno and non-Berkeley.}
Over dense-time domains, it is customary to consider only physically meaningful behaviors, namely those respecting the so-called non-Zeno property.
A behavior $b$ is non-Zeno if the sequence of transition points of $b$ has no accumulation points.
For a non-Zeno behavior $b$, it is well-defined the notions of values to the left and to the right of any transition point $t > 0$, which we denote as $b^-(t)$ and $b^+(t)$, respectively.

In this paper, we are interested in behaviors with a stronger requirement, called \emph{non-Berkeleyness}.
Informally, a behavior $b$ is non-Berkeley for some positive constant $\delta \in \reals_{> 0}$ if, for all $t \in \timedomain$, there exists a closed interval $[u, u+\delta]$ of size $\delta$ such that $t \in [u, u+\delta]$ and $b$ is constant throughout $[u, u+\delta]$.
Notice that a non-Berkeley behavior (for any $\delta$) is non-Zeno \emph{a fortiori}. 
The set of all non-Berkeley dense-time behaviors for $\delta > 0$ is denoted by $\Bchi^\delta \subset \behav_{\reals_{\geq0}}$.
In the following we always assume behaviors to be non-Berkeley, unless explicitly stated otherwise.

\paragraph{Syntax and semantics.}
From a purely semantic point of view, a (real-time) system model is simply a set of behaviors \cite{AH92b,FMMR07-TR2007-22} over some time domain $\timedomain$ and sets of items and propositions.
In practice, however, the modeler specifies a system through some suitable notation.
In this paper we consider Metric Temporal Logic (MTL) \cite{Koy90,AH93} as descriptive notation, and TA \cite{AD94,AFH96} as operational notation.
Their syntax and semantics are defined in the following.

Given an MTL formula or a TA $\mu$, and a behavior $b$, we write $b \models \mu$ to denote that $b$ describes a system evolution which satisfies all the constraints imposed by $\mu$.
If $b \models \mu$ for some $b \in \behav_{\timedomain}$, $\mu$ is called $\timedomain$-satisfiable; if $b \models \mu$ for all $b \in \behav_{\timedomain}$, $\mu$ is called $\timedomain$-valid.
Similarly, if $b \models \mu$ for some $b \in \Bchi^\delta$, $\mu$ is called $\chi^\delta$-satisfiable; if $b \models \mu$ for all $b \in \Bchi^\delta$, $\mu$ is called $\chi^\delta$-valid.

\subsection{Metric Temporal Logic}
Let $\Pcal$ be a finite (non-empty) set of atomic propositions, $\Ical$ be a finite set of items, and $\Jcal$ be the set of all (possibly unbounded) intervals of the time domain $\timedomain$ with rational endpoints.\footnote{That is any $\Ical \ni I  = \langle l, u \rangle$ for some $l \leq u$ where $l \in \timedomain \cap \rationals$ and $u \in (\timedomain \cap \rationals) \cup \{\pm\infty\}$, $\langle$ is one of $($ and $[$, and similarly for $\rangle$.}
Usually, one considers intervals with nonnegative endpoints, but we permit negative endpoints to render the presentation more uniform and straightforward.
Also, we abbreviate intervals with pseudo-arithmetic expressions, such as $=d$, $<d$, $\geq d$, for $[d,d]$, $(0,d)$, and $[d, +\infty)$, respectively.

\paragraph{MTL syntax.}
The following grammar defines the \emph{syntax} of MTL, where $I \in \Jcal$ and $\beta$ is a Boolean combination of atomic propositions or conditions over items, i.e., $\beta  ::= \pp \:|\: \iit = v \:|\: \neg \beta  \:|\:  \beta_1 \wedge \beta_2$ for $\pp \in \Pcal$, $\iit \in \Ical$, $v \in \Dcal^{\iit}$.\footnote{Note that $\neg (\iit = v)$ can be abbreviated as $\iit \neq v$.}
\begin{equation*}
  \phi  ::=  \beta \:|\: \phi_1 \vee \phi_2  \:|\: \phi_1 \wedge \phi_2  \:|\:
                   \untilMTL{I}{\beta_1, \beta_2}  \:|\: \sinceMTL{I}{\beta_1, \beta_2} \:|\:
                              \relMTL{I}{\beta_1, \beta_2}  \:|\: \redMTL{I}{\beta_1, \beta_2}
\end{equation*}

In order to ease the presentation of the discretization techniques in Section \ref{sec:discretization}, MTL formulas are introduced in a \emph{flat} normal form where negations are pushed down to (Boolean combinations of) atomic propositions, and temporal operators are not nested.
It should be clear, however, that any MTL formula can be put into this form, possibly by introducing auxiliary propositional letters \cite{DMP06,Fur07}.
The basic temporal operators of MTL are the \emph{bounded until} $\untilMTL{I}{}$ (and its past counterpart \emph{bounded since} $\sinceMTL{I}{}$), as well as its dual \emph{bounded release} $\relMTL{I}{}$ (and its past counterpart \emph{bounded trigger} $\redMTL{I}{}$).
The subscripts $I$ denote the interval of time over which every operator predicates.
In the following we assume a number of standard abbreviations, such as $\logicfalse, \logictrue, \Rightarrow, \Leftrightarrow$, and, when $I = (0, \infty)$, we drop the subscript interval of operators.
The precedence order of logic connectives is, from the one of highest binding power: $\neg, \wedge, \vee, \Rightarrow, \Leftrightarrow$.

\paragraph{MTL semantics.}
MTL \emph{semantics} is defined over behaviors, parametrically with respect to the choice of the time domain $\timedomain$.\\
\begin{tabular}{l c l}
  $b(t) \mdt \pp$ & \ \ \ iff\ \ \ &  $\pp \in b(t)|_{\Pcal}$ \\
  $b(t) \mdt \neg \pp$ & \ \ \ iff\ \ \ &  $\pp \not\in b(t)|_{\Pcal}$ \\

  $b(t) \mdt \iit = v$ & \ \ \ iff\ \ \ &  $v =  b(t)|_{\iit}$ \\
  $b(t) \mdt \iit \neq v$ & \ \ \ iff\ \ \ &  $v \neq  b(t)|_{\iit}$ \\

  $b(t) \mdt \untilMTL{I}{\beta_1, \beta_2}$ & \ \ \ iff\ \ \ &
            there exists $d \in I$ such that: $b(t+d) \mdt \beta_2$  \\
  &  &      and, for all $u \in [0, d]$ it is $b(t+u) \mdt \beta_1$  \\

  $b(t) \mdt \sinceMTL{I}{\beta_1, \beta_2}$ & \ \ \ iff\ \ \ &
            there exists $d \in I$ such that: $b(t-d) \mdt \beta_2$  \\
  &  &      and, for all $u \in [0, d]$ it is $b(t-u) \mdt \beta_1$  \\

  $b(t) \mdt \relMTL{I}{\beta_1, \beta_2}$ & \ \ \ iff\ \ \ &
            for all $d \in I$ it is: $b(t+d) \mdt \beta_2$ or there exists \\
  &  &       a $u \in [0, d)$ such that $b(t+u) \mdt \beta_1$  \\

  $b(t) \mdt \redMTL{I}{\beta_1, \beta_2}$ & \ \ \ iff\ \ \ &
            for all $d \in I$ it is: $b(t-d) \mdt \beta_2$ or there exists \\
  &  &       a $u \in [0, d)$ such that $b(t-u) \mdt \beta_1$  \\

  $b(t) \mdt \phi_1 \wedge \phi_2$  & \ \ \ iff\ \ \ & $b(t) \mdt \phi_1$ and $b(t) \mdt \phi_2$  \\

  $b(t) \mdt \phi_1 \vee \phi_2$  & \ \ \ iff\ \ \ & $b(t) \mdt \phi_1$ or $b(t) \mdt \phi_2$  \\

  $b \mdt \phi$  & \ \ \ iff\ \ \ &  for all $t \in \timedomain$: $b(t) \mdt \phi$
\end{tabular}

We remark that a global satisfiability semantics is assumed, i.e., the satisfiability of formulas is implicitly evaluated over \emph{all} time instants in the time domain.
This permits the direct and natural expression of most common real-time specifications (e.g., time-bounded response) without resorting to nesting of temporal operators.
Also notice that our MTL variant uses operators that are \emph{non-strict} in their first argument, i.e., the future and past include the present instant, and the \emph{until} and \emph{since} operators are \emph{matching}, i.e., they require their two arguments to hold together at some instant in $I$.
Other work \cite{FR07-FORMATS07} analyzes the impact of these variants on expressiveness.

\paragraph{Granularity.}
For an MTL formula $\phi$, let $\Jcal_{\phi}$ be the set of all non-null, finite interval bounds appearing in $\phi$.
Then, $\Dcal_\phi$ is the set of positive values $\delta$ such that any interval bound in $\Jcal_{\phi}$ is an integer if divided by $\delta$.

\subsubsection{\mtlplus{}/\mtlstar{} syntax and semantics.}
In order to express the discretization relations in Section \ref{sec:discretization}, it is necessary to introduce some variations of the four basic temporal operators \emph{until}, \emph{since}, \emph{release}, and \emph{trigger}, denoted as $\untilNMTL{I}{}$, $\sinceNMTL{I}{}$, $\relMMTL{I}{}$, and $\redMMTL{I}{}$, respectively.
Notice that they are not part of the language in which dense-time specifications and properties are to be expressed, and they are needed only to illustrate the discretization techniques.
We call ``\mtlplus{}'' the \emph{extension} of MTL with these operators, and ``\mtlstar{}'' the variant where we \emph{replace} the operators $\untilMTL{I}{}$, $\sinceMTL{I}{}$, $\relMTL{I}{}$, $\redMTL{I}{}$ with $\untilNMTL{I}{}$, $\sinceNMTL{I}{}$, $\relMMTL{I}{}$, and $\redMMTL{I}{}$, respectively.

Let us define the semantics of the new variants of \emph{until} and \emph{release}. \\
\begin{tabular}{l c l}
  $b(t) \mdt \untilNMTL{I}{\beta_1, \beta_2}$ & \ \ \ iff\ \ \ &
            there exists $d \in I$ such that: $b(t+d) \mdt \beta_2$  \\
  &  &      and, for all $u \in [0, d)$ it is $b(t+u) \mdt \beta_1$  \\

  $b(t) \mdt \sinceNMTL{I}{\phi_1, \phi_2}$ & \ \ \ iff\ \ \ &
            there exists $d \in I$ such that: $b(t-d) \mdt \phi_2$  \\
  &  &      and, for all $u \in [0, d)$ it is $b(t-u) \mdt \phi_1$  \\





  $b(t) \mdt \relMMTL{I}{\phi_1, \phi_2}$ & \ \ \ iff\ \ \ &
            for all $d \in I$ it is: $b(t+d) \mdt \phi_2$ or there exists \\
  &  &       a $u \in [0, d]$ such that $b(t+u) \mdt \phi_1$  \\

  $b(t) \mdt \redMMTL{I}{\phi_1, \phi_2}$ & \ \ \ iff\ \ \ &
            for all $d \in I$ it is: $b(t-d) \mdt \phi_2$ or there exists \\
  &  &       a $u \in [0, d]$ such that $b(t-u) \mdt \phi_1$
\end{tabular}

\subsubsection{Derived Temporal Operators}
It is useful to introduce a number of derived temporal operators, to be used as shorthands in writing specification formulas.
We consider those listed in Table \ref{tab:mtl-derived} ($\delta \in \reals_{> 0}$ is a parameter that will be used in the discretization technique described shortly).

\begin{table}[htb]
\begin{center}
  \begin{tabular}{|c @{$\quad \equiv \quad$} c|}
	 \hline
    \textsc{Operator}        & \textsc{Definition}  \\
    \hline
	 $\diamondMTL{I}{\beta}$    &   $\untilMTL{I}{\logictrue, \beta}$  \\
	 $\diamondPMTL{I}{\beta}$    &   $\sinceMTL{I}{\logictrue, \beta}$ \\
	 $\boxMTL{I}{\beta}$    &   $\relMTL{I}{\logicfalse, \beta}$ \\
	 $\boxPMTL{I}{\beta}$    &   $\redMTL{I}{\logicfalse, \beta}$ \\
    $\nowonstrMTL{\beta}$     &    $\untilMTL{(0, +\infty)}{\beta, \logictrue} \vee (\neg \beta \wedge \relMTL{(0, +\infty)}{\beta, \logicfalse})$  \\
    $\uptonowstrMTL{\beta}$     &    $\sinceMTL{(0, +\infty)}{\beta, \logictrue} \vee (\neg \beta \wedge \redMTL{(0, +\infty)}{\beta, \logicfalse})$ \\
	 $\nowonMTL{\beta}$  &  $\beta \wedge \nowonstrMTL{\beta}$ \\
	 $\uptonowMTL{\beta}$  &  $\beta \wedge \uptonowstrMTL{\beta}$ \\
	 $\becomesMTL{\beta_1, \beta_2}$ &   $\begin{cases}
                       \uptonowstrMTL{\beta_1} \wedge \left( \beta_2 \vee \nowonstrMTL{\beta_2} \right)   &
                                   \text{if } \timedomain = \reals_{\geq 0} \\
							  \diamondPMTL{=1}{\beta_1} \wedge \diamondMTL{[0,1]}{\beta_2}    &
                                   \text{if } \timedomain = \naturals
                                         \end{cases}$ \\
	 $\becomesOMTL{\beta_1, \beta_2}$  &  $\begin{cases}
		                 \beta_1 \wedge \diamondMTL{=\delta}{\beta_2} &
                                   \text{if } \timedomain = \reals_{\geq 0} \\
							  \beta_1 \wedge \diamondMTL{=1}{\beta_2}    &
                                   \text{if } \timedomain = \naturals
                                          \end{cases}$ \\
   \hline
  \end{tabular}
  \caption{MTL derived temporal operators}
  \label{tab:mtl-derived}
\end{center}
\end{table}

Let us describe informally the meaning of such derived operators, focusing on future ones (the meaning of the corresponding past operators is easily derivable).
$\diamondMTL{I}{\beta}$ means that $\beta$ happens within time interval $I$ in the future.
$\boxMTL{I}{\beta}$ means that $\beta$ holds throughout the whole interval $I$ in the future.
$\nowonstrMTL{\beta}$ denotes that $\beta$ holds throughout some non-empty interval in the strict future; in other words, if $t$ is the current instant, there exists some $t' > t$ such that $\beta$ holds over $(t, t')$.
Similarly, $\nowonMTL{\beta}$ denotes that $\beta$ holds throughout some non-empty interval which includes the current instant, i.e., over some $[t, t')$.
Then, $\becomesMTL{\beta_1, \beta_2}$ describes a switch from condition $\beta_1$ to condition $\beta_2$, without specifying which value holds at the current instant.
On the other hand, $\becomesOMTL{\beta_1, \beta_2}$ describes a switch from condition $\beta_1$ to condition $\beta_2$ such that $\beta_1$ holds at the current instant.

In addition, for an item $\iit$ we introduce the shorthand $\becomesMTL{\iit, v^-, v^+}$ for $\becomesMTL{\iit = v^-, \iit = v^+}$.
A similar abbreviation is assumed for $\becomesOMTL{\iit, v^-, v^+}$.

Finally, let us abbreviate by $\Alw{\phi}$ the nesting MTL formula $\phi \wedge \boxMTL{(0, +\infty)}{\phi} \wedge \boxPMTL{(0, +\infty)}{\phi}$; $b \mdt \Alw{\phi}$ iff $b \mdt \phi$, for any behavior $b$, so $\Alw{\phi}$ can be expressed without nesting if $\phi$ is flat, through the global satisfiability semantics introduced beforehand.

\subsection{Operational Model: Timed Automata} \label{sec:timedautomata}
We introduce a variant of TA which differs from the classical definitions (e.g., \cite{AD94}) in that it recognizes behaviors, rather than timed words \cite{AFH96,MNP06}.
Correspondingly, input symbols are associated with locations rather than with transitions.
Also, we introduce the following simplifications that are known to be without loss of generality: we do not define location clock invariants (also called staying conditions) and use transition guards only, and we forbid self-loop transitions.

On the other hand, we introduce one additional variant which does impact expressiveness, namely clock constraints do not distinguish between different transition edges, that is between transitions occurring right- and left-continuously.
This restriction is motivated by our ultimate goal of \emph{discretizing} TA: as it will be explained later, such distinctions would inevitably be lost in the discretization process, hence we give them up already.

Finally, for the sake of simplicity, let us not consider acceptance conditions, that is let us assume that all states are accepting.
Note, however, that introducing acceptance conditions (e.g., B\"uchi, Muller, etc.) in the formalization would be routine.

\paragraph{Timed automata syntax.}

For a set $C$ of clock variables, the set $\Phi(C)$ of \emph{clock constraints} $\xi$ is defined inductively by
\begin{displaymath}
  \xi ::= c < k  \;|\; c \geq k \;|\; \xi_1 \wedge \xi_2 \;|\; \xi_1 \vee \xi_2
\end{displaymath}
where $c$ is a clock in $C$ and $k$ is a constant in $\rationals_{\geq 0}$.

A \emph{timed automaton} $A$ is a tuple $\langle \Sigma, S, S_0, \alpha, C, E \rangle$, where:
\begin{itemize}
\item $\Sigma$ is a finite (input) alphabet,
\item $S$ is a finite set of locations,
\item $S_0 \subseteq S$ is a finite set of initial locations,
\item $\alpha: S \rightarrow 2^\Sigma$ is a location labeling function that assigns to each location $s \in S$ a set $\alpha(s)$ of propositions,
\item $C$ is a finite set of clocks, and
\item $E \subseteq S \times S \times 2^{C} \times \Phi(C)$ is a set of transitions.
  An edge $\langle s, s', \Lambda, \xi \rangle$ represents a transition from state $s$ to state $s' \neq s$; the set $\Lambda \subseteq C$ identifies the clocks to be reset with this transition, and $\xi$ is a clock constraint over $C$.
\end{itemize}

\paragraph{Timed automata semantics.}
In defining the semantics of TA over behaviors we deviate from the standard presentation (e.g., \cite{AFH96,MNP06}) in that we do not represent TA as acceptors of behaviors over the input alphabet $\Sigma$, but rather as acceptors of behaviors representing what are usually called \emph{runs} of the automaton.
In other words, we introduce automata as acceptors of behaviors over the items $\st$ and $\inpt$ representing respectively the current location and the current input symbol, as well as propositions $\rest{c}|_{c\in C}$ representing the clock reset status.
This departure from more traditional presentations is justified by the fact that we intend to provide an MTL axiomatic description of TA runs --- rather than accepted languages, which would be impossible for a well-known expressiveness gap \cite{HRS98} --- hence we define the semantics of automata over this ``extended'' state from the beginning.

Let us first define the semantics only informally.
Initially, all clocks are reset and the automaton sits in some state $s_0 \in S_0$.
At any given time $t$, when the automaton is in some state $s$, it can take nondeterministically a transition to some other state $s'$ such that $\langle s, s', \Lambda, \xi \rangle$ is a valid transition, provided the last time (before $t$) each clock has been reset is compatible with the constraint $\xi$.
If the transition is taken, all clocks in $\Lambda$ are reset, whereas all the other clocks keep on running unchanged.
Finally, as long as the automaton sits in any state $s$, the input has to satisfy the location labeling function $\alpha(s)$, namely the current input corresponds to exactly one of the propositions in $\alpha(s)$.

Formally, a timed automaton $A = \langle \Sigma, S, S_0, \alpha, C, E \rangle$ is interpreted over behaviors over items $\st: S, \inpt: \Sigma$ and propositions $R = \{\rest{c}\}_{c \in C}$.
Intuitively, at any instant of time $t$, $\st = s$ means that the automaton is in state $s$, $\inpt = \sigma$ means that the automaton is reading symbol $\sigma$, and $\rest{c}$ keeps track of resets of clock $c$ (more precisely, we model such resets through switches, from false to true or \emph{vice versa}, of $\rest{c}$).

Let $b$ be such a behavior, and let $t$ be one of its transition points.
Satisfaction of clock constraints at $t$ is defined as follows: \\
\begin{tabular}{l c l}
  $b(t) \models c < k$  & \ \ \ iff\ \ \ &  either $b^-(t) \models \rest{c}$ and there exists a $t-k < t' < t$ \\
                                       && such that $b(t') \not\models \rest{c}$; or $b^-(t) \not\models \rest{c}$ and there \\ 
                                       && exists a $t-k < t' < t$ such that $b(t') \models \rest{c}$ \\
  $b(t) \models c \geq k$  & \ \ \ iff\ \ \ &  either $b^-(t) \models \rest{c}$ and for all $t-k < t' < t:$ \\
                                       && $b'(t) \models \rest{c}$; or $b^-(t) \not\models \rest{c}$ and for all \\
                                       && $t-k < t' < t: b(t') \not\models \rest{c}$
\end{tabular} \\
Notice that this corresponds to looking for the previous time the proposition $\rest{c}$ switched (from false to true or from true to false) and counting time since then.
This requires a little hack in the definition of the semantics: namely, a first start reset of all clocks is issued before the ``real'' run begins; this is represented by time instant $\tstart$ in the formal semantics below.

Then, a behavior $b$ over $\st:S, \inpt:\Sigma, R$ (with $b: \reals_{\geq 0} \rightarrow S \times \Sigma \times 2^R$) is a \emph{run} of the automaton $A$, and we write $b \models_{\reals_{\geq 0}} A$, iff:
\begin{itemize}
\item $b(0) = \langle s_0, \sigma, \bigcup_{c \in C} \{\rest{c}\} \rangle$ and $\sigma \in \alpha(s_0)$ for some $s_0 \in S_0$;
\item there exists a transition instant $\tstart > 0$ such that: $b(t)|_{\st} = s_0$ and $b(t)|_R = R$ for all $0 \leq t \leq \tstart$, $b^-(\tstart) = \langle s_0, \sigma^-, \rho^- \rangle$ and $b^+(\tstart) = \langle s^+, \sigma^+, \rho^+ \rangle$ with $\rho^- = R$ and $\rho^+ = \emptyset$;
\item for all $t \in \reals_{\geq 0}$: $b(t)|_{\inpt} \in \alpha(b(t)|_{\st})$;
\item for all transition instants $t > \tstart$ of $b|_{\st}$ or $b|_{R}$ such that $b^-(t) = \langle s^-, \sigma^-, \rho^- \rangle$ and $b^+(t) = \langle s^+, \sigma^+, \rho^+ \rangle$, it is: $\langle s^-, s^+, \Lambda, \xi \rangle \in E$, $\sigma^- \in \alpha(s^-)$, $\sigma^+ \in \alpha(s^+)$, $\rho = \bigcup_{c \in \Lambda}\{\rest{c}\}$, $\rho^+ = \rho^- \triangle \rho = (\rho^- \setminus \rho) \cup (\rho \setminus \rho^-)$, and $b(t) \models \xi$.
\end{itemize}

\subsection{Discrete-Time Approximations of Continuous-Time \\ \ Specifications} \label{sec:discretization}
In \cite{FPR08-FM08} we presented a technique to reduce the validity problem for MTL specifications over dense time to the same problem over discrete time.
In this section we concisely summarize the fundamental results from \cite{FPR08-FM08} that are needed in the remainder of the paper, and we provide some intuition about how they can be applied to our discretization problem.

\subsubsection{Under- and Over-ap\-prox\-i\-ma\-tions of Formulas}
We introduce two approximations of MTL formulas, called under- and over-ap\-prox\-i\-ma\-tion.

\paragraph{Under-ap\-prox\-i\-ma\-tion.}
The approximation function $\underap{\cdot}$ maps dense-time MTL formulas to discrete-time \mtlstar{} formulas such that the non-validity of the latter implies the non-validity of the former, over behaviors in $\Bchi^\delta$.
More precisely, for MTL formulas such that the chosen sampling period $\delta$ is in $\Dcal_\phi$, $\underap{\cdot}$ is defined as follows.
\begin{equation*}
  \begin{array}{lcl}
  \underap{\beta}  &  \equiv  & \   \beta  \\

  \underap{\phi_1 \wedge \phi_2}  &  \equiv  & \   \underap{\phi_1} \wedge \underap{\phi_2} \\

  \underap{\phi_1 \vee \phi_2}  &  \equiv  & \  \underap{\phi_1} \vee \underap{\phi_2} \\

  \underap{\untilMTL{\langle l, u \rangle}{\phi_1, \phi_2}}  &  \equiv  &
        \  \untilNMTL{[l/\delta, u/\delta ]}{\underap{\phi_1}, \underap{\phi_2}}  \\

 \underap{\sinceMTL{\langle l, u \rangle}{\phi_1, \phi_2}}  &  \equiv  &
       \  \sinceNMTL{[l/\delta, u/\delta]}{\underap{\phi_1}, \underap{\phi_2}}  \\

  \underap{\relMTL{\langle l, u \rangle}{\phi_1, \phi_2}}  &  \equiv  &
        \  \relMMTL{\langle l/\delta , u/\delta \rangle}{\underap{\phi_1}, \underap{\phi_2}} \\

 \underap{\redMTL{\langle l, u \rangle}{\phi_1, \phi_2}}  &  \equiv  &
       \  \redMMTL{\langle l/\delta , u/\delta \rangle}{\underap{\phi_1}, \underap{\phi_2}}
  \end{array}
\end{equation*}

\paragraph{Over-ap\-prox\-i\-ma\-tion.}
The approximation function $\overap{\cdot}$ maps dense-time MTL formulas to discrete-time MTL formulas such that the validity of the latter implies the validity of the former, over behaviors in $\Bchi^\delta$.
More precisely, for MTL formulas such that the chosen sampling period $\delta$ is in $\Dcal_\phi$, $\overap{\cdot}$ is defined as follows.
\begin{equation*}
  \begin{array}{lcl}
  \overap{\beta}  &  \equiv  & \   \beta  \\

  \overap{\phi_1 \vee \phi_2}  &  \equiv  & \   \overap{\phi_1} \vee \overap{\phi_2}  \\

  \overap{\phi_1 \wedge \phi_2}  &  \equiv  & \   \overap{\phi_1} \wedge \overap{\phi_2}  \\

  \overap{\untilMTL{\langle l, u \rangle}{\phi_1, \phi_2}}
                            &  \equiv  & \  \untilMTL{[l/\delta + 1, u/\delta - 1]}{\overap{\phi_1}, \overap{\phi_2}} \\

 \overap{\sinceMTL{\langle l, u \rangle}{\phi_1, \phi_2}}
                           &  \equiv  & \  \sinceMTL{[l/\delta + 1, u/\delta - 1]}{\overap{\phi_1}, \overap{\phi_2}} \\

  \overap{\relMTL{\langle l, u \rangle}{\phi_1, \phi_2}}
                            &  \equiv  & \  \relMTL{[l/\delta - 1, u/\delta + 1]}{\overap{\phi_1}, \overap{\phi_2}} \\

 \overap{\redMTL{\langle l, u \rangle}{\phi_1, \phi_2}}
                           &  \equiv  & \  \redMTL{[l/\delta - 1, u/\delta + 1]}{\overap{\phi_1}, \overap{\phi_2}}
  \end{array}
\end{equation*}

\subsubsection{System Verification through Approximation}
We have the following fundamental verification result from \cite{FPR08-FM08}, which provides a justification for the TA verification technique discussed in this paper.

\begin{proposition}[Approximations \cite{FPR08-FM08}] \label{prop:approximations}
  For any MTL formulas $\phi_1, \phi_2$, and for any $\delta \in \Dcal_{\phi_1, \phi_2}$:
  (1) if $\Alw{{\underap{\phi_1}}} \Rightarrow \Alw{\overap{\phi_2}}$ is $\naturals$-valid,
		    then $\Alw{\phi_1} \Rightarrow \Alw{\phi_2}$ is $\chi^\delta$-valid;
and (2) if $\Alw{\overap{\phi_1}} \Rightarrow \Alw{\underap{\phi_2}}$ is not $\naturals$-valid,
		    then $\Alw{\phi_1} \Rightarrow \Alw{\phi_2}$ is not $\chi^\delta$-valid.
\end{proposition}

\subsubsection{Discussion}
Proposition \ref{prop:approximations} suggests a verification technique which builds two formulas through a suitable composition of over- and under-ap\-prox\-i\-ma\-tions of the system description and the putative properties, and it infers the validity of the properties from the results of a discrete-time validity checking.
The technique is incomplete as, in particular, when approximation (1) is not valid and approximation (2) is valid we cannot infer anything about the validity of the property in the original system over dense time.

Let us now provide some evidence about why different, but equivalent, dense-time formulas can yield dramatically different --- in terms of usefulness --- approximated discrete-time formulas.
We provide one in-the-small example for over-ap\-prox\-i\-ma\-tions and one for under-ap\-prox\-i\-ma\-tions.
More concrete examples will appear in Section \ref{sec:TA-approx} when building approximations of TA's axiomatic description.

Let us consider dense-time MTL formula $\theta_1 = \boxMTL{(0,\delta)}{\pp}$ which, under the global satisfiability semantics, says that $\pp$ is \emph{always} true.
Its under-ap\-prox\-i\-ma\-tion is $\underap{\theta_1} = \boxMTL{\emptyset}{\pp}$ which holds for any discrete-time behavior!
Thus, we have an under-ap\-prox\-i\-ma\-tion which is likely too coarse, as it basically adds no information to the discrete-time representation.
So, if we build formula (1) from Proposition \ref{prop:approximations} with $\underap{\theta_1}$ in it, it is most likely that the antecedent will be trivially satisfiable (because $\underap{\theta_1}$ introduces no constraint) and hence formula (1) will be non-valid, yielding no information to the verification process.
If, however, we modify $\theta_1$ into the \emph{equivalent} $\theta_1' = \pp \wedge \theta_1$ we get an under-ap\-prox\-i\-ma\-tion which can be written as simply $\underap{\theta_1'} = \pp$, which correctly entails that $\pp$ is always true over discrete-time as well.
This is likely a much better approximation, one which better preserves the original ``meaning'' of $\theta_1$.

Let us now consider dense-time MTL formula $\theta_2 = \diamondMTL{[0,2\delta]}{\pp}$, which describes a proposition $\pp$ which is false for no longer than $2\delta$ time units.
If we compute its over-ap\-prox\-i\-ma\-tion, we get $\overap{\theta_2} = \diamondMTL{=1}{\pp}$ which, under the global satisfiability semantics, entails that $\pp$ is always true.
Although the actual assessment depends on the role $\theta$ plays in the overall specification, it is likely that this over-ap\-prox\-i\-ma\-tion is too coarse, as it basically adds ``too strong'' information to the discrete-time representation.
So, if we build formula (2) from Proposition \ref{prop:approximations} with $\overap{\theta_2}$ in it, it is very likely that the antecedent will be unsatisfiable (because $\overap{\theta_2}$ introduces a very strong constraint) and hence formula (2) will be valid, yielding no information to the verification process.
On the contrary, if we simply modify $\theta_2$ into the \emph{equivalent} $\theta_2' = \pp \vee \theta_2$ we get an over-ap\-prox\-i\-ma\-tion which can be written as $\overap{\theta_2'} = \diamondMTL{[0,1]}{\pp}$, i.e., $\pp$ is false no more than every two time steps.
This looks like a much better approximation, one which better preserves the original ``meaning'' of $\theta_2$.

\section{Formalizing Timed Automata in MTL} \label{sec:formalizingTAs}
Let us consider a timed automaton $A = \langle \Sigma, S, S_0, \alpha, C, E \rangle$ and let us formalize its runs over non-Berkeley behaviors for some $\delta > 0$.
In other words, we are going to provide a set of formulas $\phi_1, \ldots, \phi_{\textup{\ref{ax:liveness}}}$ such that, for all non-Berkeley behaviors $b$, $b \models A$ iff $b \models \phi_j$ for all $j = 1, \ldots, \ref{ax:liveness}$.

\paragraph{Translating clock constraints.}
We associate an MTL formula $\Xi(\xi)$ to every clock constraint $\xi$ such that $b(t) \models \xi$ iff $b(t) \models \Xi(\xi)$ at all transition points $t$.
$\Xi(\xi)$ can be defined inductively as:
\begin{displaymath}
  \begin{array}{lcl}
	 \Xi\left(c < k\right)      &   \equiv  &
              \uptonowstrMTL{\rest{c}} \wedge \diamondPMTL{(0, k)}{\neg \rest{c}}
               \quad \vee \quad \uptonowstrMTL{\neg \rest{c}} \wedge \diamondPMTL{(0, k)}{\rest{c}}\\
	 \Xi\left(c \geq k\right)      &   \equiv  &
              \uptonowstrMTL{\rest{c}} \wedge \boxPMTL{(0, k)}{\rest{c}}
              \quad \vee \quad \uptonowstrMTL{\neg \rest{c}} \wedge \boxPMTL{(0, k)}{\neg \rest{c}} \\
	 \xi_1 \wedge \xi_2  &   \equiv &  \Xi_1 \wedge \Xi_2 \\
	 \xi_1 \vee \xi_2  &   \equiv  &  \Xi_1 \vee \Xi_2
  \end{array}
\end{displaymath}
Basically, $\Xi$ translates the guard $\xi$ by comparing the current time to the last time a reset for the clock $c$ happened, where a reset is signaled by a switching of item $\rest{c}$.
Notice that this assumes the existence of a ``first reset'' of all clocks, as specified in the formal semantics of TA, and as will be postulated in Formula \frf{ax:start} below.
Also notice that, when computing the approximations of the clock-constraint formulas, we will have to require that every constant $k$ used in the definition of the TA is an integral multiple of $\delta$.

\paragraph{Necessary conditions for state change.}
Let us state the necessary conditions that characterize a state change.
For any pair of states $s_i, s_j \in S$ such that there are $K$ transitions $\langle s_i, s_j, \Lambda^k, \xi^k \rangle \in E$ for all $1 \leq k \leq K$, we introduce the axiom:
\begin{multline} \label{ax:si2sj}
  \becomesMTL{\st, s_i, s_j} \;\Rightarrow\; 
     \bigvee_k  \Xi(\xi^k) \wedge \bigwedge_{c \in \Lambda^k}
               \Big( \becomesMTL{\neg \rest{c}, \rest{c}} \vee \becomesMTL{\rest{c}, \neg \rest{c}} \Big)
\end{multline}

Complementarily, we introduce an axiom to assert that for any pair of states $s_i \neq s_j \in S$ such that $\langle s_i, s_j, \Lambda, \xi \rangle \not\in E$ for any $\sigma, \Lambda, \xi$, i.e., for any pair of states that are not connected by any edge:
\begin{equation} \label{ax:si2sjforbidden}
  \neg \becomesMTL{\st, s_i, s_j}
\end{equation}

\paragraph{Sufficient conditions for state change.}
We have multiple sufficient conditions for state changes; basically, they account for reactions to reading input symbols and resetting clocks.
Let us consider input first: the staying condition in every state must be satisfied always, so for all $s \in S$ we add the axiom:
\begin{equation} \label{ax:invariance}
  \st = s \qquad \Rightarrow \qquad \inpt \in \alpha(s)
\end{equation}

Then, for each reset of a clock $c \in C$, let us consider all edges of the form $\langle s_i^k, s_j^k, \Lambda^k, \xi^k \rangle \in E$, such that $c \in \Lambda^k$.
Hence, we introduce the pair of axioms:
\begin{gather} 
  \becomesMTL{\neg \rest{c}, \rest{c}} \quad \Rightarrow \quad 
       \bigvee_k \becomesMTL{\st, s_i^k, s_j^k} \nonumber \\
  \becomesMTL{\rest{c}, \neg \rest{c}} \quad \Rightarrow \quad 
       \bigvee_k \becomesMTL{\st, s_i^k, s_j^k}  \vee \bigvee_{s_0 \in S_0} \boxPMTL{(0, +\infty)}{\bigwedge_{c \in C} \rest{c} \wedge \st = s_0} \label{ax:restc}
\end{gather}
Note that the second axiom has an additional part that takes into account the instants before the first reset (which must occur somewhere as shown in \frf{ax:start}, and which corresponds to the instants before $\tstart$ in the formal semantics), whereas the first one is not applicable before such a first reset.

\paragraph{Initialization and liveness condition.}
We complete our axiomatization by first describing the system initialization.

We remark that the following axiom is only evaluated at $0$.
Notice that, under the global satisfiability semantics and with a mono-infinite time domain, a formula $\phi^0$ that should be only evaluated at $0$ can be expressed as $\boxPMTL{}{\logicfalse{}} \Rightarrow \phi^0$, as $\boxPMTL{}{\logicfalse{}}$ holds only where there is no past, i.e., at $0$.
\begin{equation} \label{ax:start}
  \text{at $0$:}\quad \bigwedge_{c \in C} \rest{c} \wedge \diamondMTL{[0,2\delta]}{\bigwedge_{c \in C}\neg\rest{c}}
                      \wedge \bigvee_{s_0 \in S_0} \nowonMTL{\st = s_0}
\end{equation}
Notice that we make the axiomatization slightly more ``deterministic'' than the formal semantics, in that we require that $\tstart$, when the first reset of the clocks occurs, is between $0$ and $2\delta$; this, combined with the non-Berkeleyness requirement, says that it actually occurs between $\delta$ and $2\delta$.
All in all, \frf{ax:start} pictures the following initialization:
\begin{itemize}
  \item $\rest{c}$ holds over $[0,\delta]$ for all $c \in C$;
  \item $\rest{c}$ switches to false at some $\tstart \in (\delta, 2\delta]$ for all $c \in C$ (clearly, this transition point is the same for all $c \in C$, still because of the non-Berkeleyness assumption);
  \item $\st = s_0$ holds for some $s_0 \in S_0$ over $[0,\delta]$;
  \item because of the non-Berkeleyness assumption, if $\st$ changes in $(\delta, 2\delta]$ it does so together with the resets at $\tstart$;
  \item $\becomesMTL{\rest{c}, \neg \rest{c}}$ holds at $\tstart$ for all $c \in C$; the consequent of \frf{ax:restc} is true because of the disjunct $\boxPMTL{(0, +\infty)}{\bigwedge_{c \in C} \rest{c} \wedge \st = s_0}$ which holds at $\tstart$.
\end{itemize}

Finally, often we introduce a ``liveness'' condition which states that we eventually have to move out of every state, corresponding to the fact that all states are accepting \emph{\`a la} B\"uchi.
Thus, for every state $s \in S$, let $S'_s \subset S$ be the set of states that are directly reachable from $s$ through a single transition; then we consider the axiom:
\begin{equation} \label{ax:liveness}
  \st = s \;\Rightarrow\; \diamondMTL{}{\bigvee_{s' \in S'_s} \st = s'}
\end{equation}

 \subsection{About the Correctness and Completeness of the Axiomatization}
 We omit a proof of the completeness and correctness of the axiomatization; we refer the reader to \cite[App.~D.6]{Fur07} where a proof for a similar axiomatization is sketched.
Here, we just add a few remarks that can help justify the correctness and appropriateness of the present axiomatization.

\begin{proposition}[MTL TA Axiomatization]
Let $A = \langle \Sigma, S, S_0, \alpha, C, E \rangle$ be a timed automaton, $\phi_1^A, \ldots, \phi_{\textup{\ref{ax:liveness}}}^A$ be formulas \fsrf{ax:si2sj}{ax:liveness} for TA A, and let $b \in \Bchi^\delta$ be any non-Berkeley behavior over items $\st: S, \inpt: \Sigma$ and propositions in $R$.
Then $b \models A$ for some $\tstart\in(\delta, 2\delta)$\footnote{This additional condition is introduced to take into account the particular form of the initialization axiom \frf{ax:start}.} if and only if $b \models \bigwedge_{1 \leq j \leq \textup{\ref{ax:liveness}}} \phi_j^A$.
\end{proposition}

\paragraph{State changes can occur right- or left-continuously.}
It should be clear that the above axiomatization with the \emph{becomes} operators does not force any item to transition either right- or left-continuously; in fact, the operator allows both possibilities.
Over dense time, however, it would have been possible to force transitions to occur either always right- or always left-continuously.
For instance, right-continuity can be achieved in one of the following ways:
\begin{itemize}
  \item add formulas such as $\nowonstrMTL{\st = s_i} \Rightarrow s_i$;
  \item add formulas such as $\neg \left( \uptonowMTL{\st = s_i} \wedge \nowonstrMTL{\st = s_j} \right)$.
\end{itemize}
Correspondingly, the whole formalization could have been simplified a bit taking into account this new property.

Unfortunately, however, it is not difficult to see that all solutions would yield very poor discrete-time over-ap\-prox\-i\-ma\-tions, where by very poor we mean comprising only very trivial behaviors, and thus offering a very weak support to verification.
For instance, the over- and under-ap\-prox\-i\-ma\-tions of $\nowonMTL{\st = s_i} \Rightarrow s_i$ would require $\st$ to stay equal to $s_i$ forever once it takes such value.
Intuitively, this is due to the fact that a fine-grained information such as the edge of items at transition points is lost with a finite-precision sampling.
There may be work-arounds for this, but it seems that they are overly complex.
On the other hand, forgetting about characterizing transitions as right- or left-continuous allows us to get a much more straightforward axiomatization while still getting our approximations to work reasonably well.

\section{Discrete-Time Approximations of Timed Automata} \label{sec:TA-approx}
Let us show how to compute the under- and over-ap\-prox\-i\-ma\-tion of formulas \fsrf{ax:si2sj}{ax:liveness} in a suitable way.

\subsection{Under-ap\-prox\-i\-ma\-tion}

The particular form of formulas \fsrf{ax:si2sj}{ax:si2sjforbidden},\frf{ax:restc} is unsuitable to produce under-ap\-prox\-i\-ma\-tions that are strong enough to be useful.

Let us first of all notice that $\underap{\nowonstrMTL{\beta}} = \diamondMTL{[0,1]}{\beta}$ and $\underap{\uptonowstrMTL{\beta}} = \diamondPMTL{[0,1]}{\beta}$.
In	fact, over dense time, the definition of the \emph{nowon} operator can be rewritten equivalently as: $\beta \wedge \untilMTL{(0, +\infty)}{\beta, \logictrue} \vee \neg \beta \wedge \relMTL{(0, +\infty)}{\beta, \logicfalse}$, whose under-ap\-prox\-i\-ma\-tion is: $\beta \wedge \untilNMTL{ }{\beta, \logictrue} \vee \neg \beta \wedge \relMMTL{ }{\beta, \logicfalse}$.
	 Over discrete time, the latter is equivalent to $\beta \vee \neg \beta \wedge \diamondMTL{=1}{\beta} = \diamondMTL{[0,1]}{\beta}$. 
Correspondingly, $\underap{\becomesMTL{\beta_1, \beta_2}} = \diamondMTL{[0,1]}{\beta_1} \wedge \diamondMTL{[0,1]}{\beta_2}$.
Then, for $\beta_1,\beta_2$ that cannot hold at the same instant (i.e., $\neg(\beta_1 \wedge \beta_2)$), this approximation is a suitable discrete-time representation of a transition from $\beta_1$ to $\beta_2$.
However, consider $\underap{\neg \becomesMTL{\beta_1, \beta_2}} = \underap{\uptonowstrMTL{\neg \beta_1} \vee \neg \beta_2 \wedge \nowonstrMTL{\neg \beta_2}} = \diamondPMTL{[0,1]}{\neg \beta_1} \vee \neg \beta_2 \wedge \diamondMTL{[0,1]}{\neg \beta_2} = \diamondPMTL{[0,1]}{\neg \beta_1} \vee \neg \beta_2 = \neg (\boxPMTL{[0,1]}{\beta_1} \wedge \beta_2)$.
There are two problems with this result.
First, $\underap{\neg \becomesMTL{\beta_1, \beta_2}} \neq \neg \underap{\becomesMTL{\beta_1, \beta_2}}$; since we use $\becomesMTL{\beta_1, \beta_2}$ to describe transitions, there are discrete-time behaviors where such a transition both occurs and does not occur, i.e., $\underap{\becomesMTL{\beta_1, \beta_2}}$ and $\underap{\neg \becomesMTL{\beta_1, \beta_2}}$ are both true.
Second, $\underap{\neg \becomesMTL{\beta_1, \beta_2}}$ is very weak, in that it is true, in particular, whenever $\beta_1$ or $\beta_2$ are false; since $\becomesMTL{\beta_1, \beta_2}$ is often used as antecedent of implications in our axiomatization, such implications are trivially true because $\neg \beta_1 \vee \neg \beta_2$ is an identity when $\beta_1,\beta_2$ cannot hold at the same instant.

This demands a thorough revision of the axiomatization, in order to make it amenable to under-approximations.

\subsubsection{A New Axiomatization}
The new axiomatization basically replaces every occurrence of $\becomesMTL{\beta_1, \beta_2}$ with $\becomesLMTL{\beta_1, \beta_2}$.
Hence, formulas \fsrf{ax:si2sj}{ax:si2sjforbidden},\frf{ax:restc} are changed as follows (notice that also $\Xi$ is changed into $\XiL$, as we are explaining shortly).

\begin{multline} \label{ax:si2sjua}
  \becomesLMTL{\st, s_i, s_j} \;\Rightarrow\; 
     \bigvee_k  \XiL(\xi^k) \wedge \bigwedge_{c \in \Lambda^k}
               \Big( \becomesLMTL{\neg \rest{c}, \rest{c}} \vee \becomesLMTL{\rest{c}, \neg \rest{c}} \Big)
\end{multline}

\begin{equation} \label{ax:si2sjforbiddenua}
  \neg \becomesLMTL{\st, s_i, s_j}  
\end{equation}

\begin{gather} 
  \becomesLMTL{\neg \rest{c}, \rest{c}} \quad \Rightarrow \quad 
       \bigvee_k \becomesLMTL{\st, s_i^k, s_j^k} \nonumber \\
  \becomesLMTL{\rest{c}, \neg \rest{c}} \quad \Rightarrow \quad 
       \bigvee_k \becomesLMTL{\st, s_i^k, s_j^k}  \vee \bigvee_{s_0 \in S_0} \boxPMTL{[0, +\infty)}{\rest{c} \wedge \st = s_0} \label{ax:restcua}
\end{gather}

\begin{displaymath}
  \begin{array}{lcl}
	 \XiL\left(c < k\right)      &   \equiv  &
              \rest{c} \wedge \diamondPMTL{(0, k)}{\neg \rest{c}}
               \quad \vee \quad \neg \rest{c} \wedge \diamondPMTL{(0, k)}{\rest{c}}\\
	 \XiL\left(c \geq k\right)      &   \equiv  &
              \rest{c} \wedge \boxPMTL{(0, k-\delta)}{\rest{c}}
              \quad \vee \quad \neg \rest{c} \wedge \boxPMTL{(0, k-\delta)}{\neg \rest{c}} \\
  \end{array}
\end{displaymath}

Let us now show that the new axiomatization --- where formulas \fsrf{ax:si2sj}{ax:si2sjforbidden},\frf{ax:restc} are replaced by the new formulas \fsrf{ax:si2sjua}{ax:restcua} --- is indeed equivalent to the old one.

\begin{proof}[\textbf{Proof that \frf{ax:si2sj} iff \frf{ax:si2sjua}.}]
Let us first show that \frf{ax:si2sj} implies \frf{ax:si2sjua}, so let $t$ be the current instant, assume that \frf{ax:si2sj} and the antecedent $\becomesLMTL{\st, s_i, s_j}$ of \frf{ax:si2sjua} hold: we establish that the consequent of \frf{ax:si2sjua} holds.
$\becomesLMTL{\st, s_i, s_j}$ means that $\st = s_i$ at $t$ and $\st = s_j \neq s_i$ at $t+\delta$; hence there must be a transition instant $t'$ of item $\st$ somewhere in $[t,t+\delta]$.
Then \frf{ax:si2sj} evaluated at $t'$ entails that $t'$ is a transition instant for some propositions $\rest{c}|_{c \in \Lambda^k}$ as well.
Let $d \in C$ be anyone of such clocks and assume that $\becomesMTL{\rest{d}, \neg \rest{d}}$ holds at $t'$.
Let us first assume $t' \in (t,t+\delta)$; correspondingly, from the non-Berkeleyness assumption, $\rest{d}$ holds over $[t,t')$ and $\neg \rest{d}$ holds over $(t', t+\delta]$.
In particular, $\rest{d}$ holds at $t$ and $\neg \rest{d}$ holds at $t+\delta$, so $\becomesLMTL{\rest{d}, \neg \rest{d}}$ holds at $t$.
Otherwise, let $t' = t$, so $\st$ changes its value left-continuously at $t$.
Then, again from \frf{ax:si2sj} and the non-Berkeleyness assumption, $\rest{d}$ also changes its value left-continuously, so $\rest{d}$ holds at $t$ and $\neg \rest{d}$ holds at $t+\delta$.
Finally, if $t' = t+\delta$, $\st$ changes its value right-continuously at $t'$, so $\rest{d}$ also changes its value right-continuously, so $\rest{d}$ holds at $t$ and $\neg \rest{d}$ holds at $t+\delta$.
In all, since $d$ is generic, and the same reasoning applies for the converse transition $\becomesMTL{\neg \rest{d}, \rest{d}}$, we have established that $\bigwedge_{c \in \Lambda^k} \left(\becomesLMTL{\neg \rest{c}, \rest{c}} \vee \becomesLMTL{\rest{c}, \neg \rest{c}}\right)$ holds at $t$. \\
Next, let us establish $\XiL(\xi^k)$ from $\Xi(\xi^k)$.
Let us first consider some $\Xi\left(d < k\right)$ such that $\uptonowstrMTL{\rest{d}} \wedge \diamondPMTL{(0, k)}{\neg \rest{d}}$ at $t'$.
So, let $t'' \in (t'-k, t')$ be the largest instant with a transition from $\neg \rest{d}$ to $\rest{d}$.
Note that it must actually be $t'' \in (t'-k, t]$ because $t'-t \leq \delta$ and the non-Berkeleyness assumption.
If $t'' \in (t'-k, t) \subseteq (t-k, t)$ then $\rest{d} \wedge \diamondPMTL{(0,k)}{\neg \rest{d}}$ holds at $t$, hence $\XiL\left(d < k\right)$ is established.
If $t'' = t$ then $\rest{d}$ switches to true right-continuously at $t$, so $\rest{d} \wedge \uptonowstrMTL{\neg \rest{d}}$ at $t$ which also entails $\XiL\left(d < k\right)$.
The same reasoning applies if $\uptonowstrMTL{\neg \rest{c}} \wedge \diamondPMTL{(0, k)}{\rest{c}}$ holds at $t'$.
Finally, consider some $\Xi\left(d \geq k\right)$ such that $\uptonowstrMTL{\rest{d}} \wedge \boxPMTL{(0, k)}{\rest{d}}$ holds at $t$, thus $\rest{d}$ holds over $(t-k, t)$.
From $t \leq t'+\delta$ we have $t'+\delta-k \geq t+k$ so $(t'-k+\delta, t') \subseteq (t-k, t)$, which shows that $\boxPMTL{(0, k-\delta)}{\rest{d}}$ holds at $t'$.
The usual reasoning about transition edges would allow us to establish that also $\rest{d}$ holds at $t'$.
Since the same reasoning applies if $\uptonowstrMTL{\neg \rest{d}} \wedge \boxPMTL{(0, k)}{\neg \rest{d}}$, we have established that $\XiL(d \geq k)$ holds at $t'$.
Since $d$ is generic, we have that $\XiL(\xi^k)$ holds at $t'$.

Let us now prove \frf{ax:si2sjua} implies \frf{ax:si2sj}, so let $t$ be the current instant, assume that \frf{ax:si2sjua} and the antecedent $\becomesMTL{\st, s_i, s_j}$ of \frf{ax:si2sj} hold: we establish that the consequent of \frf{ax:si2sj} holds.
So, there is a transition of $\st$ from $s_i$ to $s_j \neq s_i$ at $t$; from the non-Berkeleyness assumption we have that $\st = s_i$ and $\st = s_j$ hold over $[t-\delta, t)$ and $(t, t+\delta]$, respectively.
If the transition of $\st$ is left-continuous (i.e., $\st = s_i$ holds at $t$), consider \frf{ax:si2sjua} at $t$, where the antecedent holds.
So, $\XiL(\xi^k) \wedge \bigwedge_{c \in \Lambda^k} \left( \becomesLMTL{\neg \rest{c}, \rest{c}} \vee \becomesLMTL{\rest{c}, \neg \rest{c}} \right)$ holds at $t$ for some $k$.
Let $d \in \Lambda^k$ be such that $\becomesLMTL{\neg \rest{d}, \rest{d}}$ holds, that is $\neg \rest{d}$ holds at $t$ and $\rest{d}$ holds at $t+\delta$.
This entails that there exists a transition point $t' \in [t, t+\delta]$ of $\rest{d}$.
However, $t$ is already a transition point, thus it must be $t' = t$; this shows $\becomesMTL{\neg \rest{d}, \rest{d}}$ at $d$.
Recall that $d$ is generic, and the same reasoning applies for the converse transition from $\rest{d}$ to $\neg \rest{d}$.
If, instead, the transition of $\st$ is right-continuous (i.e., $\st = s_j$ holds at $t$), we consider \frf{ax:si2sjua} at $t-\delta$ and perform a similar reasoning.
All in all, we have established that $\bigwedge_{c \in \Lambda^k} \left( \becomesMTL{\neg \rest{c}, \rest{c}} \vee \becomesMTL{\rest{c}, \neg \rest{c}} \right)$ holds at $t$.\\
The clock constraint formula $\Xi(\xi^k)$ can also be proved along the same lines.
For instance, assume that the transition of $\st$ at $t$ is left-continuous and $\uptonowMTL{\rest{d}}$ holds at $t$ for some $d \in C$, and consider a constraint $\XiL(d < k)$ at $t$.
We have that $\diamondPMTL{(0,k)}{\neg \rest{d}}$ must holds at $t$, which establishes that $\Xi(d < k)$ holds at $t$.
Similar reasonings apply to the other cases.
\end{proof}

\begin{proof}[\textbf{Proof that \frf{ax:si2sjforbidden} iff \frf{ax:si2sjforbiddenua}.}]
Let $\becomesMTL{\st, s_i, s_j}$ holds at $t$; we prove that $\becomesLMTL{\st, s_i, s_j}$ at some $t'$.
If the transition of $\st$ at $t$ is right-continuous let $t' = t+\delta$, else let $t' = t$.
From the non-Berkeleyness assumption we have that $\st = s_j$ at $t+\delta$ and $\st = s_i$ at $t-\delta$.
Correspondingly, $\becomesLMTL{\st, s_i, s_j}$ holds at $t'$ because $\st = s_i$ at $t'$ and $\st = s_j$ at $t'+\delta$.

For the converse, let $\becomesLMTL{\st, s_i, s_j}$ holds at $t$; we prove that $\becomesMTL{\st, s_i, s_j}$ at some $t'$.
This is immediate because $\st = s_i$ at $t$ and $\st = s_j$ at $t+\delta$ entail that there exists a transition instant $t' \in [t, t+\delta]$ where $\becomesMTL{\st, s_i, s_j}$ holds.
\end{proof}

\begin{proof}[\textbf{Proof that \frf{ax:restc} iff \frf{ax:restcua}.}]
The proof of this part is along the same lines as for the proof that \frf{ax:si2sj} iff \frf{ax:si2sjua}.
\end{proof}

In the following sub-sections we are going to compute under-approximations of these new equivalent axiomatization, thus showing that the results are indeed much more satisfactory than with the original axioms.
In fact, we can already see that $\underap{\becomesLMTL{\beta_1, \beta_2}} = \beta_1 \wedge \diamondMTL{=1}{\beta_2} = \neg (\neg \beta_1 \vee \diamondMTL{=1}{\neg \beta_2}) = \neg \underap{\neg \becomesLMTL{\beta_1, \beta_2}}$, thus solving the fundamental problem with the previous axiomatization.

\subsubsection{Clock Constraints}
Let us consider the under-ap\-prox\-i\-ma\-tions of clock constraints; they are both straightforward.

\begin{equation*}
  \begin{array}{lcl}
	 \underap{\XiL\left(c < k\right)}      &   \equiv  &
	           \rest{c} \wedge \diamondPMTL{[0,k/\delta]}{\neg \rest{c}}
				  \quad \vee \quad
	           \neg \rest{c} \wedge \diamondPMTL{[0,k/\delta]}{\rest{c}} \\
	 \underap{\XiL\left(c \geq k\right)}      &   \equiv  &
	           \rest{c} \wedge \boxPMTL{[1,k/\delta - 2]}{\rest{c}}
				  \quad \vee \quad
	           \neg \rest{c} \wedge \boxPMTL{[0,k/\delta - 2]}{\neg \rest{c}}
  \end{array}
\end{equation*}

\subsubsection{Formulas \fsrf{ax:si2sj}{ax:si2sjforbidden}}
From the preliminaries, it is straightforward to re-write \frf{ax:si2sjua} in normal form, compute the under approximation, and re-write the resulting discrete-time formula as:
\begin{multline} \label{ax:si2sj-underap}
  \becomesLMTL{\st, s_i, s_j} \;\Rightarrow\; 
     \bigvee_k  \underap{\XiL(\xi^k)} \wedge \bigwedge_{c \in \Lambda^k}
               \Big( \becomesLMTL{\neg \rest{c}, \rest{c}} \vee \becomesLMTL{\rest{c}, \neg \rest{c}} \Big)
\end{multline}

The under-ap\-prox\-i\-ma\-tion of \frf{ax:si2sjforbiddenua} is also straightforward:
\begin{equation} \label{ax:si2sjforbidden-underap}
  \neg \becomesLMTL{\st, s_i, s_j}
\end{equation}

\subsubsection{Formulas \fsrf{ax:invariance}{ax:restc}}
Formula \frf{ax:restcua} has a structure similar to formula \frf{ax:si2sjua}; so we immediately compute its under-ap\-prox\-i\-ma\-tions as:
\begin{gather} 
  \becomesLMTL{\neg \rest{c}, \rest{c}} \quad \Rightarrow \quad 
       \bigvee_k \becomesLMTL{\st, s_i^k, s_j^k} \nonumber \\
  \becomesLMTL{\rest{c}, \neg \rest{c}} \quad \Rightarrow \quad 
       \bigvee_k \becomesLMTL{\st, s_i^k, s_j^k}  \vee \bigvee_{s_0 \in S_0} \boxPMTL{[0, +\infty)}{\rest{c} \wedge \st = s_0} \label{ax:restc-underap}
\end{gather}

Also, simply $\underap{\frf{ax:invariance}} = \frf{ax:invariance}$.

\subsubsection{Formulas \fsrf{ax:start}{ax:liveness}}
Formula \frf{ax:liveness} is unchanged under under-approximation (after noticing that \linebreak $\diamondMTL{[0, \infty)}{\phi}$ is equivalent to $\diamondMTL{}{\phi}$ when the antecedent of \frf{ax:liveness} holds), so $\underap{\frf{ax:liveness}} = \frf{ax:liveness}$.
Formulas \fsrf{ax:start}{ax:liveness} are straightforward to under-approximate, and they produce discrete-time formulas that are perfectly adequate.

Next, let us consider \frf{ax:start} instead.
Since $\underap{\neg \boxPMTL{}{\logicfalse}} = \logictrue$, we first re-write it as:
\begin{equation} \label{ax:startua}
\boxPMTL{[\delta, +\infty]}{\logicfalse} \Rightarrow
\bigwedge_{c \in C} \rest{c} \wedge \diamondMTL{[0,2\delta]}{\bigwedge_{c \in C}\neg\rest{c}}
                      \wedge \bigvee_{s_0 \in S_0} \st = s_0
\end{equation}
Let us discuss why \frf{ax:startua} and \frf{ax:start} are equivalent, when considered together with the other axioms.
$\boxPMTL{[\delta, +\infty]}{\logicfalse}$ holds precisely over $[0, \delta)$, thus \frf{ax:startua} asserts that:
\begin{itemize}
  \item $\rest{c}$ holds over $[0,\delta)$ for all $c \in C$;
  \item $\rest{c}$ switch to false at some $\tstart \in [\delta, 2\delta]$ for all $c \in C$;
  \item $\st = s_0$ holds for some $s_0 \in S_0$ over $[0,\delta)$;
	 note that is must be the same $s_0$ throughout the interval, still because of the non-Berkeley assumption;
  \item because of the non-Berkeleyness assumption, if $\st$ changes in $(\delta, 2\delta]$ it does so together with the resets at $\tstart$;
  \item $\becomesLMTL{\rest{c}, \neg \rest{c}}$ holds over $[\tstart-\delta,\tstart)$ for all $c \in C$; the consequent of \frf{ax:restcua} is true because of the disjunct $\boxPMTL{[0, +\infty)}{\bigwedge_{c \in C} \rest{c} \wedge \st = s_0}$ which holds throughout $[0,\tstart)$.
\end{itemize}
All in all the new initialization formula forces a behavior which is the same as in the original one.
Then, given that $\underap{\neg \boxPMTL{[\delta, +\infty)}{\logicfalse}} = \diamondMTL{}{\logictrue}$ which holds everywhere except at $0$, we compute $\underap{\frf{ax:startua}}$:
\begin{equation} \label{ax:start-underap}
  \text{at $0$:}\quad \bigwedge_{c \in C} \rest{c} \wedge \diamondMTL{[1,2]}{\bigwedge_{c \in C}\neg\rest{c}}
                      \wedge \bigvee_{s_0 \in S_0} \st = s_0
\end{equation}
where $\diamondMTL{[0,2]}{\bigwedge_{c \in C}\neg\rest{c}}$ has been rewritten as $\diamondMTL{[1,2]}{\bigwedge_{c \in C}\neg\rest{c}}$ because $\bigwedge_{c \in C} \rest{c}$ holds at $0$.

In addition, we notice the following fact.
Assume that $\bigwedge_{c \in C}\neg\rest{c}$ holds at $1$; then \frf{ax:restc} can require a state transition only for instants $\geq 1$.
Otherwise, assume that $\bigwedge_{c \in C}\neg\rest{c}$ holds at $2$ at that \emph{some} resets switch at $1$, i.e., there exists a $D \subset C$ such that: (a) $\bigwedge_{c \in C} \rest{c}$ at $0$, (b) $\bigwedge_{c \in D} \rest{c}$ at $1$, and (c) $\bigwedge_{c \in C} \neg \rest{c}$ at $2$.
Then, \frf{ax:restc} requires a state transition at $1$.
All in all, \frf{ax:restc-underap} can be rewritten equivalently without the $\bigvee_{s_0 \in S_0} \boxPMTL{[0, +\infty)}{\rest{c} \wedge \st = s_0}$ part if it is evaluated only at instants $\geq 1$.

\subsection{Over-ap\-prox\-i\-ma\-tion}
Formulas \fsrf{ax:si2sj}{ax:liveness} are in a form which is unsuitable to compute useful over-ap\-prox\-i\-ma\-tion.
Hence, we follow the same path as for the under-ap\-prox\-i\-ma\-tion: we introduce a different, albeit equivalent, continuous-time axiomatization, which is then amenable to over-ap\-prox\-i\-ma\-tion.

\subsubsection{Preliminaries} \label{sec:prelim}
Let us consider a generic Boolean combination $\beta$ and let us compute the following over-ap\-prox\-i\-ma\-tions (clearly, the justifications for those with past operators are the same as for the future operators, so they are omitted for brevity):
\begin{itemize}
  \item $\overap{\nowonstrMTL{\beta}} = \boxMTL{[0,1]}{\beta}$. \\
	 From the definition of the \emph{nowon} operator, we have: $\untilMTL{\geq 1}{\beta, \logictrue} \vee (\neg \beta \wedge \relMTL{\geq -1}{\beta, \logicfalse})$.
	 Over discrete time, it is easy to check that $\untilMTL{\geq 1}{\beta, \logictrue}$ is equivalent to $\boxMTL{[0, 1]}{\beta}$; on the other hand, the second disjunct $\neg \beta \wedge \relMTL{\geq -1}{\beta, \logicfalse}$ is equivalent to $\logicfalse$, as when $d \leq 0$ the interval $[0, d)$ is empty.

  \item $\overap{\diamondMTL{[0,2\delta]}{\beta}} = \diamondMTL{=1}{\beta}$.

  \item $\overap{\nowonMTL{\beta}} = \boxMTL{[0,1]}{\beta}$.
	 
  \item $\overap{\uptonowstrMTL{\beta}} = \overap{\uptonowMTL{\beta}} = \boxPMTL{[0,1]}{\beta}$.

  \item $\overap{\neg \becomesMTL{\beta_1, \beta_2}} = \neg (\becomesMTL{\beta_1, \beta_2} \vee \becomesOMTL{\beta_1, \beta_2})$, assuming $\beta_1,\beta_2$ cannot hold at the same instant. \\
	 Recall the definition of $\becomesMTL{\beta_1, \beta_2}$, so $\neg \becomesMTL{\beta_1, \beta_2} = \uptonowstrMTL{\neg \beta_1} \vee (\neg \beta_2 \wedge \nowonstrMTL{\neg \beta_2})$.
	 Thus, $\overap{\neg \becomesMTL{\beta_1, \beta_2}} = \boxPMTL{[0,1]}{\neg \beta_1} \vee (\neg \beta_2 \wedge \boxMTL{[0,1]}{\neg \beta_2}) = \boxPMTL{[0,1]}{\neg \beta_1} \vee \boxMTL{[0,1]}{\neg \beta_2}$.
	 By pushing negations outward in the latter, we get: $\neg ( \diamondPMTL{[0,1]}{\beta_1} \wedge \diamondMTL{[0,1]}{\beta_2} )$, which is equivalent to $\neg \left( \becomesMTL{\beta_1, \beta_2} \vee \becomesOMTL{\beta_1, \beta_2} \right)$ if $\beta_1,\beta_2$ cannot hold at the same instant.
\end{itemize}

\subsubsection{Clock Constraints}
It is not difficult to compute the over-ap\-prox\-i\-ma\-tions of the ``existential'' clock constraint.
In fact, we have:
\begin{displaymath}
	 \overap{\Xi\left(c < k\right)}      \;\equiv\;
              \boxPMTL{[0,1]}{\rest{c}} \wedge \diamondPMTL{[1, k/\delta-1]}{\neg \rest{c}}
              \;\vee\; \boxPMTL{[0,1]}{\neg \rest{c}} \wedge \diamondPMTL{[1, k/\delta-1]}{\rest{c}}
\end{displaymath}

On the contrary, we have to ``massage'' the ``universal'' clock constraints into a more suitable form; otherwise, e.g., $\overap{\boxPMTL{(0,k)}{\rest{c}}} =  \boxPMTL{[-1,k/\delta+1]}{\rest{c}}$ but the latter is never satisfiable if $c$ is both checked and reset when a transition is taken.
We can, however, perform a transformation where $\Xi(c \geq k)$ becomes:
\begin{displaymath}
  \Xi\left(c \geq k\right)      \quad\equiv\quad
              \uptonowstrMTL{\rest{c}} \wedge \boxPMTL{[\delta, k)}{\rest{c}}
              \quad\vee\quad  \uptonowstrMTL{\neg \rest{c}} \wedge \boxPMTL{[\delta, k)}{\neg \rest{c}}
\end{displaymath}
which is seen to be equivalent for non-Berkeley behaviors at transition points (when clock constraints are evaluated).
Hence, we have:
\begin{displaymath}
	 \overap{\Xi\left(c \geq k\right)}      \;\equiv\;
              \boxPMTL{[0,1]}{\rest{c}} \wedge \boxPMTL{[0, k/\delta+1]}{\rest{c}}
              \;\vee\;  \boxPMTL{[0,1]}{\neg \rest{c}} \wedge \boxPMTL{[0, k/\delta+1]}{\neg \rest{c}}
\end{displaymath}

\subsubsection{Attempting Formula \frf{ax:si2sj}} \label{sec:attempt}
It is not difficult to see that formula \frf{ax:si2sj} yields a very poor over-ap\-prox\-i\-ma\-tion.
In particular, the portions in the consequent corresponding to the clock resets: $\becomesMTL{\neg \rest{c}, \rest{c}} \vee \becomesMTL{\rest{c}, \neg \rest{c}}$ become, when over-approximated:
\begin{multline}
  \overap{ \uptonowstrMTL{\rest{c}} \wedge (\neg \rest{c} \vee \nowonstrMTL{\neg \rest{c}})
           \;\;\vee\;\;
			  \uptonowstrMTL{\neg \rest{c}} \wedge (\rest{c} \vee \nowonstrMTL{\rest{c}})} \;=\\
         \boxPMTL{[0,1]}{\rest{c}} \wedge (\neg \rest{c} \vee \boxMTL{[0,1]}{\neg \rest{c}})
          \;\;\vee\;\;
			 \boxPMTL{[0,1]}{\neg \rest{c}} \wedge (\rest{c} \vee \boxMTL{[0,1]}{\rest{c}})
\end{multline}

Clearly, the above discrete-time formula is unsatisfiable, as, for instance, $\boxPMTL{[0,1]}{\rest{c}}$ is in contradiction with $\neg \rest{c} \vee \boxMTL{[0,1]}{\neg \rest{c}}$.
Similar problems arise with the over-ap\-prox\-i\-ma\-tions of formula \frf{ax:restc}.

As a consequence, the over-ap\-prox\-i\-ma\-tion axioms would only be satisfiable with behaviors where the antecedents are identically false.
It is not difficult to realize that such behaviors would be the trivial ones, where no transition ever happens.
This in turn would contradict (the over-ap\-prox\-i\-ma\-tion of) formula \frf{ax:liveness}.
So, overall, we end up with a set of over-approximated axioms which are unsatisfiable; clearly, this is of little interest for checking non-validity, as an unsatisfiable set of axioms entails any property.

\subsubsection{A New Axiomatization}
However, we can rewrite our axioms in a form which is equivalent but which yields much better discrete-time over-ap\-prox\-i\-ma\-tions.

Let us rewrite formulas \frf{ax:si2sj},\frf{ax:restc} as follows (formulas \fsrf{ax:si2sjforbidden}{ax:invariance} are instead unchanged).

\begin{multline} \label{ax:si2sjoa}
  \becomesMTL{\st, s_i, s_j} \;\Rightarrow\;
     \bigvee_k 
     \left( \begin{array}{c}
         \Xi(\xi^k) \wedge
         \bigwedge_{c \in \Lambda^k} \left( \begin{array}{c}
             \uptonowstrMTL{\neg \rest{c}} \wedge \boxMTL{=\delta}{\st = s_j \Rightarrow \rest{c}} \\
             \vee \\ 
             \uptonowstrMTL{\rest{c}} \wedge \boxMTL{=\delta}{\st = s_j \Rightarrow \neg \rest{c}}
             \end{array} \right) 
          \end{array} \right)
\end{multline}

\begin{gather}
  \becomesMTL{\neg \rest{c}, \rest{c}} \quad \Rightarrow \quad 
    \bigvee_k \left( \uptonowstrMTL{\st = s_i^k} \wedge \boxMTL{=\delta}{\rest{c} \Rightarrow \st = s_j^k} \right)
    \nonumber \\
  \becomesMTL{\rest{c}, \neg \rest{c}} \quad \Rightarrow \quad 
       \bigvee_k \left( \uptonowstrMTL{\st = s_i^k} \wedge \boxMTL{=\delta}{\neg \rest{c} \Rightarrow \st = s_j^k} \right)
		 \nonumber \\
       \qquad \qquad \qquad \qquad \vee \bigvee_{s_0 \in S_0} \boxPMTL{[\delta, +\infty)}{\rest{c} \wedge \st = s_0}
       \label{ax:restcoa}
\end{gather}

We claim that these new axioms describe the \emph{same} behaviors as the original axioms \fsrf{ax:si2sj}{ax:restc}.

\begin{proof}[\textbf{Proof that \frf{ax:si2sj} iff \frf{ax:si2sjoa}.}]
Since the antecedents of \frf{ax:si2sj} and \frf{ax:si2sjoa} are the same, we just have to prove that the consequents are equivalent, assuming that the antecedents hold.
So let $\becomesMTL{\st, s_i, s_j}$ hold at the current instant $t$; this means that item $\st$ transitions from $s_i$ to $s_j$.
In particular, notice that the non-Berkeleyness requirement for $\delta$ entails that $s_j$ holds at least over the interval $(t, t+\delta]$.

Now, let $d \in C$.
Note that $\nowonstrMTL{\rest{d}}$ at $t$ iff $\st = s_j \Rightarrow \rest{d}$ at $t+\delta$, because $t$ is a transition point, so the non-Berkeleyness requirement entails that $\rest{d}$ holds throughout $(t, t+\delta]$.
Hence, $\becomesMTL{\neg \rest{d}, \rest{d}}$ iff $\uptonowstrMTL{\neg \rest{d}} \wedge \boxMTL{=\delta}{\st = s_j \Rightarrow \rest{d}}$, at $t$.
Since the reasoning holds for a generic clock, and also for the converse transition from $\rest{d}$ to $\neg \rest{d}$, and $\Xi(\xi^k)$ is the same in both \frf{ax:si2sj} and \frf{ax:si2sjoa} we have proved that \frf{ax:si2sj} iff \frf{ax:si2sjoa}.
\end{proof}

\begin{proof}[\textbf{Proof that \frf{ax:restc} iff \frf{ax:restcoa}.}]
Proofs along the very same lines can be provided for formulas formulas \frf{ax:restc} and \frf{ax:restcoa}.
We only notice that the term $\boxPMTL{(0, +\infty)}{\rest{c} \wedge \st = s_0}$ has been equivalently changed to $\boxPMTL{[\delta, +\infty)}{\rest{c} \wedge \st = s_0}$.
In fact, \frf{ax:start} entails that $\st = s_0$ holds throughout $[0,\delta]$, hence $\becomesMTL{\rest{c}, \neg \rest{c}}$ is false over $[0,\delta)$.
We omit all other details for brevity.
\end{proof}

\subsubsection{Formulas \fsrf{ax:si2sj}{ax:si2sjforbidden}}
The newly built formula \frf{ax:si2sjoa} is now amenable to over-ap\-prox\-i\-ma\-tion.
In fact, we have the following discrete-time formula.
\begin{multline} \label{ax:si2sj-overap}
   \becomesMTL{\st, s_i, s_j} \vee  \becomesOMTL{\st, s_i, s_j} 
   \quad \Rightarrow \\ \bigvee_k \left( \begin{array}{c}
       \overap{\Xi(\xi^k)} \wedge 
         \bigwedge_{c \in \Lambda^k} \left( \begin{array}{c}
             \boxPMTL{[0,1]}{\neg \rest{c}} \wedge \boxMTL{[0,2]}{\st = s_j \Rightarrow \rest{c}} \\
             \vee \\
             \boxPMTL{[0,1]}{\rest{c}} \wedge \boxMTL{[0,2]}{\st = s_j \Rightarrow \neg \rest{c}}
             \end{array} \right)
          \end{array} \right)
\end{multline}

Notice instead that the over-ap\-prox\-i\-ma\-tion of \frf{ax:si2sjforbidden} is simply:
\begin{equation} \label{ax:si2sjforbidden-overap}
  \neg ( \becomesMTL{\st, s_i, s_j} \vee \becomesOMTL{\st, s_i, s_j})
\end{equation}

\subsubsection{Formula \frf{ax:restc}}
Formula \frf{ax:restcoa} has a structure similar to formula \frf{ax:si2sjoa}; so we immediately compute its over-ap\-prox\-i\-ma\-tions as:
\begin{gather}
  \becomesMTL{\neg \rest{c}, \rest{c}} \vee \becomesOMTL{\neg \rest{c}, \rest{c}} \quad \Rightarrow \quad 
    \bigvee_k \left( \boxPMTL{[0,1]}{\st = s_i^k} \wedge \boxMTL{[0,2]}{\rest{c} \Rightarrow \st = s_j^k} \right)
    \nonumber \\
  \becomesMTL{\rest{c}, \neg \rest{c}} \vee \becomesOMTL{\rest{c}, \neg \rest{c}} \quad \Rightarrow \quad 
    \bigvee_k \left( \boxPMTL{[0,1]}{\st = s_i^k} \wedge \boxMTL{[0,2]}{\neg \rest{c} \Rightarrow \st = s_j^k} \right) \nonumber \\
    \qquad \qquad \qquad \qquad \qquad \qquad \vee \bigvee_{s_0 \in S_0} \boxPMTL{[0, +\infty)}{\rest{c} \wedge \st = s_0}
       \label{ax:restc-overap}
\end{gather}

\subsubsection{Some Simplifications}
In this section we show how to re-write discrete-time formulas \fsrf{ax:si2sj-overap}{ax:restc-overap} above in a simpler but equivalent form.

Let us start by noting that the formulas have a similar structure, and in particular have antecedents that are structurally identical, the only difference being the items they predicate about.
In fact, these antecedent describe a transition of an item from a value to another value; so \frf{ax:si2sj-overap} describes a transition of item $\st$ from $s_i$ to $s_j$, \frf{ax:restc-overap} a transition of some $\rest{c}$, etc.

Let us consider a generic current instant $h$ where the antecedent of \frf{ax:si2sj-overap} holds and let us spell out what form the transition of $\st$ can take.
$\becomesMTL{\st, s_i, s_j} \vee \becomesOMTL{\st, s_i, s_j}$ holds precisely in the following three cases:
\begin{enumerate}
  \item \label{cs2} $\st = s_i$ holds at $h-1$ and $\st = s_j$ holds at $h$;
  \item \label{cs1} $\st = s_i$ holds at $h-1$ and $\st = s_j$ holds at $h+1$;
  \item \label{cs3} $\st = s_i$ holds at $h$ and $\st = s_j$ holds at $h+1$.
\end{enumerate}
We are going to show that case \ref{cs2} is in contradiction with the other axioms, and therefore can be removed from the axiomatization.

So, assume that $\st = s_i$ holds at $h-1$ and $\st = s_j$ holds at $h$.
The consequent of \frf{ax:si2sj-overap} is then contradictory: $\boxPMTL{[0,1]}{\neg \rest{c}}$ implies that $\rest{c}$ is false at $h$, but $\boxMTL{[0,2]}{\st = s_j \Rightarrow \rest{c}}$ implies that $\rest{c}$ is true at $h$ because $\st = s_j$ is the case.
All similarly if $\boxPMTL{[0,1]}{\rest{c}}$ holds.

It is simple to see that similar contradictions arise if we consider a transition for $\rest{c}$ from false to true (or true to false) for some $c \in C$.
We conclude that we should never consider transitions as in case \ref{cs2}.

Now, notice that if case \ref{cs2} never holds, case \ref{cs1} reduces to case \ref{cs3}.
In fact, it cannot be $\st = s_j$ at $h$ or we would have case \ref{cs2}, so it must be $\st = s_i$ at $h$.
All in all, every antecedent in formulas \fsrf{ax:si2sj-overap}{ax:restc-overap} can be simplified into just $\becomesOMTL{\st, s_i, s_j} \equiv \st = s_i \wedge \diamondMTL{=1}{\st = s_j}$ and similar ones.

Finally, notice that also formula \frf{ax:si2sjforbidden-overap} can be simplified into just $\neg \becomesOMTL{\st, s_i, s_j}$.
To see this, assume to the contrary that $\st = s_i$ holds at $h-1$ and $\st = s_j$ holds at $h$, for some pair of states $s_i,s_j$ which do not belong to any transition.
In this case, $\becomesOMTL{\st, s_i, s_j}$ holds at $h - 1$, thus the new formula is false, which shows that such a transition cannot occur even with the new, weaker formula.

All in all, we have formulas \fsrf{ax:si2sj-overap}{ax:restc-overap} simplified as follows.
\begin{multline} \label{ax:si2sj-overap-simp}
   \becomesOMTL{\st, s_i, s_j} \quad \Rightarrow \\
       \bigvee_k \left( \begin{array}{c}
           \overap{\Xi(\xi^k)} \wedge 
         \bigwedge_{c \in \Lambda^k} \left( \begin{array}{c}
             \boxPMTL{[0,1]}{\neg \rest{c}} \wedge \boxMTL{[0,2]}{\st = s_j \Rightarrow \rest{c}} \\
             \vee \\
             \boxPMTL{[0,1]}{\rest{c}} \wedge \boxMTL{[0,2]}{\st = s_j \Rightarrow \neg \rest{c}}
             \end{array} \right)
             \end{array} \right)
\end{multline}
\begin{equation} \label{ax:si2sjforbidden-overap-simp}
  \neg \becomesOMTL{\st, s_i, s_j}
\end{equation}

\begin{gather}
  \becomesOMTL{\neg \rest{c}, \rest{c}} \quad \Rightarrow \quad 
    \bigvee_k \left( \boxPMTL{[0,1]}{\st = s_i^k} \wedge \boxMTL{[0,2]}{\rest{c} \Rightarrow \st = s_j^k} \right)
    \nonumber \\
  \becomesOMTL{\rest{c}, \neg \rest{c}} \quad \Rightarrow \quad 
    \bigvee_k \left( \boxPMTL{[0,1]}{\st = s_i^k} \wedge \boxMTL{[0,2]}{\neg \rest{c} \Rightarrow \st = s_j^k} \right)
	 \nonumber \\
    \qquad \qquad \qquad \vee \bigvee_{s_0 \in S_0} \boxPMTL{[0, +\infty)}{\rest{c} \wedge \st = s_0}
       \label{ax:restc-overap-simp}
\end{gather}

\subsubsection{Formulas \frf{ax:invariance},\fsrf{ax:start}{ax:liveness}}
Notice that simply $\overap{\frf{ax:invariance}} = \frf{ax:invariance}$ and $\overap{\frf{ax:liveness}} = \frf{ax:liveness}$.

For \frf{ax:start} notice that $\overap{\neg \boxPMTL{}{\logicfalse}} = \diamondPMTL{[1,+\infty)}{\logictrue}$ which holds everywhere except at $0$.
Thus, we can write $\overap{\frf{ax:start}}$ as:
\begin{equation} \label{ax:start-overap}
  \text{at $0$:}\quad \bigwedge_{c \in C}\rest{c} \wedge \diamondMTL{=1}{\bigwedge_{c \in C}\neg \rest{c}}
                \wedge \bigvee_{s_0 \in S_0} \boxMTL{[0,1]}{ \st = s_0 }
\end{equation}
Notice that \frf{ax:start-overap} entails that $\boxPMTL{[0, +\infty)}{\rest{c} \wedge \st = s_0}$ holds for some $s_0 \in S_0$ at $0$.
Correspondingly, \frf{ax:restc-overap} can be rewritten equivalently without the $\bigvee_{s_0 \in S_0} \boxPMTL{[0, +\infty)}{\rest{c} \wedge \st = s_0}$ part if it is evaluated only at instants $\geq 1$.

\subsection{Summary}
The following proposition summarizes the results of the discrete-time approximation formulas.

\begin{proposition} \label{prop:summary}
Let $S$ be a real-time system described by timed automaton $A = \langle \Sigma, S, S_0, \alpha, C, E \rangle$ and by a set of MTL specification formulas $\{\system_j\}_{j}$ over items in $\IIT$ and propositions in $\Pcal$.
Also, let $\prop$ be another MTL formula over items in $\IIT \cup \{\st:S, \inpt:\Sigma\}$ and propositions in $\Pcal \cup R$.
Then:
\begin{itemize}
\item if:
 \begin{multline*}
   \Alw{\phi^A_{\textup{\frf{ax:si2sj-underap}}} \wedge \phi^A_{\textup{\frf{ax:si2sjforbidden-underap}}} \wedge \phi^A_{\textup{\frf{ax:invariance}}} \wedge \phi^A_{\textup{\frf{ax:restc-underap}}} \wedge \phi^A_{\textup{\frf{ax:start-underap}}} \wedge \phi^A_{\textup{\frf{ax:liveness}}}  \wedge \bigwedge_{j} \underap{\system_j}}
   \\ \Rightarrow \Alw{\overap{\prop}}
\end{multline*}
is $\naturals$-valid, then $\prop$ is satisfied by all non-Berkeley runs $b \in \Bchi^\delta$ of the system (with $\tstart \in (\delta, 2\delta)$);

\item if:
 \begin{multline*}
   \Alw{\phi^A_{\textup{\frf{ax:si2sj-overap-simp}}} \wedge \phi^A_{\textup{\frf{ax:si2sjforbidden-overap-simp}}} \wedge \phi^A_{\textup{\frf{ax:invariance}}} \phi^A_{\textup{\frf{ax:restc-overap-simp}}} \wedge \phi^A_{\textup{\frf{ax:start-overap}}} \wedge \phi^A_{\textup{\frf{ax:liveness}}} \wedge \bigwedge_{j} \overap{\system_j}}
   \\ \Rightarrow \Alw{\underap{\prop}}
\end{multline*}
is not $\naturals$-valid, then $\prop$ is not satisfied by all non-Berkeley runs $b \in \Bchi^\delta$ of the system (with $\tstart \in (\delta, 2\delta)$).
\end{itemize}
\end{proposition}

\section{Implementation and Example} \label{sec:impl-example}
This section describes briefly the implementation of the verification technique introduced in the previous section and it discusses an example of system verified with the resulting tool.

\subsection{\tazot{}}
We implemented the verification technique of this paper as a plugin to the \zot{} bounded satisfiability checker \cite{zot,PMS07} named \tazot{}.
The plugin provides a set of primitives by which the user can provide the description of a timed automaton, of a set of MTL axioms, and a set of MTL properties (to be verified).
The tool then automatically builds the two discrete-time approximation formulas of Proposition \ref{prop:summary}.
These are checked for validity over time $\naturals$ bounded by some user-defined constant; the results of the validity check allows one to infer the validity of the original dense-time models, according to Proposition \ref{prop:summary}.

More precisely, the verification process in \tazot{} consists of three sequential phases.
First, the discrete-time MTL formulas of Proposition \ref{prop:summary} are built and are translated into a propositional satisfiability (SAT) problem.
Second, the SAT instance is put into conjunctive normal form (CNF), a standard input format for SAT solvers.
Third, the CNF formula is fed to a SAT solving engine (such as MiniSat, zChaff, or MiraXT) for the validity checking.

\subsection{A Communication Protocol Example}
We demonstrate the practical feasibility of our verification techniques by means of an example, where we verify certain properties of a communication protocol, modeled through a timed automaton.

\subsubsection{Description of the Protocol}
Let us consider a server accepting requests from clients to perform a certain service (the exact nature of the service is irrelevant for our purposes).
Initially, the server is \emph{idle} in a passive open state.
At any time, a client can initiate a protocol run; when this is the case, the server moves to a \emph{try} state.
Within $T_1$ time units, the state moves to a new \emph{s$_1$} state, characterizing the first request of the client for the service.
The request can either terminate within $T_2$ time units, or time-out after $T_2$ time units have elapsed.
When it terminates, it can do so either successfully (\emph{ok}) or unsuccessfully (\emph{ko}).
In case of success, the protocol run is completed afterward, and the server goes back to being \emph{idle}.
In case of failure or time-out, the server moves to a new \emph{s$_2$} state for a second attempt.
The second attempt is executed all similarly to the first one, with the only exception that the system goes back to the \emph{idle} state afterward, regardless of the outcome (success, failure, or time-out).

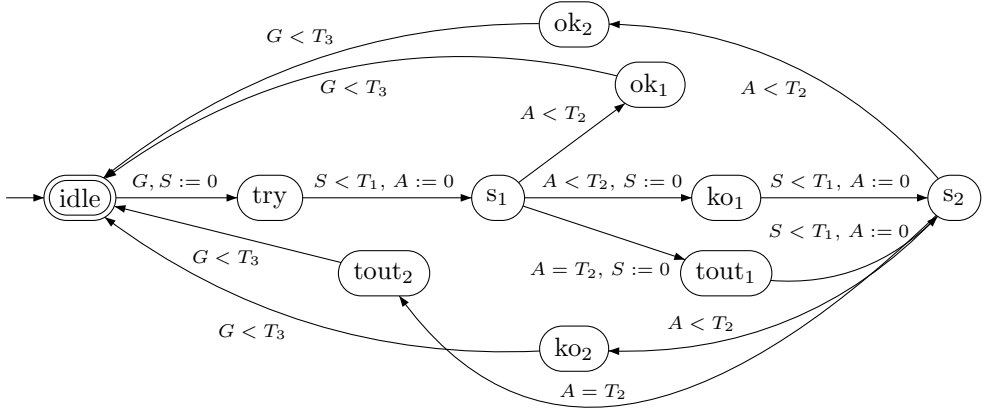
\begin{figure}
\begin{center}
\begin{picture}(125,58)(-25,-50)
  \gasset{Nadjust=w,Nadjustdist=2,Nh=6,Nmr=3}
  \node[Nmarks=ir](A)(-20,-20){idle}
  \node(B)(5,-20){try}
  \node(C)(35,-20){s$_1$}
  \node(D)(55,-5){ok$_1$}
  \node(E)(65,-20){ko$_1$}
  \node(F)(65,-30){tout$_1$}
  \node(G)(95,-20){s$_2$}
  \node(H)(45,3){ok$_2$}
  \node(I)(45,-40){ko$_2$}
  \node(J)(20,-30){tout$_2$}
  \drawedge(A,B){\scriptsize $G,S := 0$}
  \drawedge(B,C){\scriptsize $S < T_1$, $A:=0$}
  \drawedge(C,D){\scriptsize $A < T_2$}
  \drawedge(C,E){\scriptsize $A < T_2$, $S:=0$}
  \drawedge[ELside=r](C,F){\scriptsize $A = T_2$, $S:=0$}
  \drawedge(E,G){\scriptsize $S < T_1$, $A:=0$}
  \drawedge[ELpos=50,ELdist=1.5,ELside=l,curvedepth=-5](F,G){\scriptsize $S < T_1$, $A:=0$}
  \drawedge[curvedepth=-7](G,H){\scriptsize $A < T_2$}
  \drawedge[ELpos=70,ELside=r,curvedepth=7](G,I){\scriptsize $A < T_2$}
  \drawqbedge[ELpos=55,ELside=r](G,45,-70,J){\scriptsize $A = T_2$}
  \drawedge[curvedepth=-10](D,A){\scriptsize $G < T_3$}
  \drawedge[curvedepth=-7,ELside=r](H,A){\scriptsize $G < T_3$}
  \drawedge[ELpos=60,curvedepth=7](I,A){\scriptsize $G < T_3$}
  \drawedge(J,A){\scriptsize $G < T_3$}
\end{picture}
\end{center}
\caption{Timed automaton modeling the communication protocol.} \label{fig:trans_prot}
\end{figure}

The timed automaton of Figure \ref{fig:trans_prot} models the protocol.
Recall that the definition of clock constraints given in Section \ref{sec:timedautomata} forbids the introduction of exact constraints such as $A = T_2$.
Hence, we mean clock constraints in the form $C = T$ as a shorthand for the valid clock constraint $T \leq C < T + \delta$, where $\delta$ is the chosen sampling period.
In other words, we approximate exact clock constraints to within a tolerance which is given by the time granularity $\delta$.

\subsubsection{Properties of the System}
Let us describe the properties we verified using our technique.
We verified 5 properties of a single instance of the automaton, and 2 other properties of a concurrent run of two (or more) instances of the automaton, synchronized according to additional MTL axioms described below.
We included a \emph{false} property among the former 5, in order to show how the verification technique works at disproving false properties.

\paragraph{Single instance properties.}
\begin{enumerate}
\item \label{p1}
  ``If there is a success, the server goes back to idle without passing through error states.''
\begin{equation*}
 \ok \vee \okk \quad\Rightarrow\quad  \untilMTL{}{\ko \vee \koo ,\idle}
\end{equation*}

\item \label{p2}
  ``If there is a failure, the server goes back to idle without passing through success states.''
\begin{equation*}
 \ko \vee \koo \quad\Rightarrow\quad  \untilMTL{}{\ok \vee \okk ,\idle}
\end{equation*}
   This property is false, and in fact counterexamples are produced in the tests.

\item \label{p3}
  ``A full run of the protocol executes in no more than $T_3$ time units.''
\begin{equation*}
 \try \quad\Rightarrow\quad  \diamondMTL{(0,T_3)}{\idle}
\end{equation*}
  This property, as it is, falls in the incompleteness area of the method.
  In fact, whether a run is completed in $T_3/\delta$ time instants depends sensibly on how the sampling is chosen, so the method cannot conclude anything within its accuracy.
  However, if we slightly weaken the property by changing $T_3$ into $T_3 + \delta$ the method is successful in verifying the property.
  In the tables, the (verified) property --- modified in this way --- is labeled \ref{p3}'.

\item \label{p4}
  ``The first attempt of the protocol is initiated no later than $2T_1 + T_2 + \delta$ time units after the run has been initiated.''
\begin{equation*}
 \s \quad\Rightarrow\quad  \diamondPMTL{(0,2T_1 + T_2 + \delta)}{\try}
\end{equation*}

\item \label{p5}
  ``A run is terminated within $T_3$ time units after a successful outcome, without going through failure states.''
\begin{equation*}
 \ok \quad\Rightarrow\quad  \untilMTL{(0,T_3)}{\neg(\ko \vee \koo), \idle}
\end{equation*}
\end{enumerate}

\paragraph{Concurrent run properties.}
Let us now assume that the server runs two concurrent instances of the same protocol.
Since the two processes run on the same hardware, it is reasonable to assume that the outcomes of two parallel protocol runs will be correlated.
More precisely, we assume that two parallel protocol runs that are initiated concurrently either both terminate successfully, or both terminate unsuccessfully.
To formalize this assumption, we augment our operational model with the following MTL axiom, where corresponding states of the two automata instances are differentiated by a superscripted $A$ or $B$:
\begin{multline} \label{eq:work-together}
 \try^A \wedge \try^B \Rightarrow \\
\untilMTL{}{\neg(\toutt^A \vee \koo^A), \ok^A \vee \okk^A} \wedge \untilMTL{}{\neg(\toutt^B \vee \koo^B), \ok^B \vee \okk^B} \\
\vee \\
 \untilMTL{}{\neg(\ok^A \vee \okk^A), \toutt^A \vee \koo^A} \wedge \untilMTL{}{\neg(\ok^B \vee \okk^B), \toutt^B \vee \koo^B}
\end{multline}

It is also simple to conceive a generalization of \frf{eq:work-together} to $N \geq 2$ concurrent runs, where we re-state the same property for every pair of instances, that is:
\begin{multline} \label{eq:work-togeter-multiple}
\forall 1 \leq i < j \leq N:
 \quad \try^i \wedge \try^j \Rightarrow \\
\untilMTL{}{\neg(\toutt^i \vee \koo^i), \ok^i \vee \okk^i} \wedge \untilMTL{}{\neg(\toutt^j \vee \koo^j), \ok^j \vee \okk^j} \\
\vee \\
 \untilMTL{}{\neg(\ok^i \vee \okk^i), \toutt^i \vee \koo^i} \wedge \untilMTL{}{\neg(\ok^j \vee \okk^j), \toutt^j \vee \koo^j}
\end{multline}

Correspondingly, we introduce the following two properties to be verified in this concurrent system.
\begin{enumerate}
\setcounter{enumi}{5}
\item \label{pc1}
  ``If at some time one process succeeds and the other fails, then they have not begun the current run together.''
\begin{equation*}
 \okk^A \wedge \koo^B \quad\Rightarrow\quad  \sinceMTL{(0,T_3)}{\neg(\try^A \wedge \try^B),\try^A \vee \try^B}
\end{equation*}

\item \label{pc2}
  ``If at some time one process succeeds and the other failed recently, then they have not begun the current run together.''
\begin{equation*}
 \okk^A \wedge \diamondPMTL{(0,T_1)}{\koo^B} \quad\Rightarrow\quad  \sinceMTL{(0,T_3)}{\neg(\try^A \wedge \try^B),\try^A \vee \try^B}
\end{equation*}
\end{enumerate}

\subsection{Experimental Evaluation}
Tables \ref{tab:single} shows some results obtained in tests with \tazot{} verifying the properties above.
In all tests it is $\delta = 1$.
For each test the table reports: the checked property; the number $N_r$ of parallel protocol runs, according to which the discretizations are built; the values of other parameters in the model (i.e., $T_1, T_2, T_3$); the size $k$ of the explored state space (as \zot{} is a bounded satisfiability checker); the total amount of time and space (in MBytes) to perform each phase of the verification, namely formula building (\textsc{FB}), transformation into conjunctive normal form (\textsc{CNF}), and propositional satisfiability checking (\textsc{SAT}); and the total size (in thousands of clauses) of the propositional formulas that have been checked.

The tests have been performed on a PC equipped with an AMD Athlon64 X2 Dual Core Processor 4000+, 2 Gb of RAM, and Kubuntu GNU/Linux (kernel 2.6.22).
\tazot{} used GNU CLisp v.~2.41 and MiniSat v.~2.0 as SAT-solving engine.

\begin{table}[htbp]
\begin{center}
\begin{scriptsize}
\begin{tabular}{|c r r r l l l l|}
\hline
\textsc{Pr.\#} & $N_r$ & $T_1,T_2,T_3$  &  $k$  &  \textsc{FB} (time/mem)  &  \textsc{CNF} (time/mem) & \textsc{SAT} (time/mem) &  \textsc{\# KCl.} \\
\hline
1  &  1  &  3,6,18  &  30  &  0.1 min/114.6 Mb  &  3.9 min  &  0.3 min/90.2 Mb & 520.2 \\
2  &  1  &  3,6,18  &  30  &  0.1 min/228.6 Mb  &  7.8 min  &  0.5 min/180.1 Mb & 1037.9 \\
3  &  1  &  3,6,18  &  30  &  0.2 min/244.3 Mb  &  9.1 min  &  0.7 min/195.6 Mb & 1112.4 \\
3'  &  1  &  3,6,18  &  30  &  0.1 min/122.5 Mb  &  4.6 min  &  0.4 min/98.0 Mb & 557.7 \\
4  &  1  &  3,6,18  &  30  &  0.1 min/121.4 Mb  &  4.5 min  &  0.3 min/97.4 Mb & 553.2 \\
5  &  1  &  3,6,18  &  30  &  0.1 min/122.6 Mb  &  4.6 min  &  0.4 min/97.9 Mb & 557.3 \\
\hline
1  &  1  &  3,6,24  &  36  &  0.1 min/146.8 Mb  &  6.3 min  &  0.5 min/117.9 Mb & 669.1 \\
2  &  1  &  3,6,24  &  36  &  0.2 min/292.9 Mb  &  12.5 min  &  0.9 min/235.4 Mb & 1335.2 \\
3  &  1  &  3,6,24  &  36  &  0.2 min/319.0 Mb  &  15.4 min  &  1.2 min/258.6 Mb & 1459.0 \\
3'  &  1  &  3,6,24  &  36  &  0.1 min/159.9 Mb  &  7.6 min  &  0.7 min/129.3 Mb & 731.3 \\
4  &  1  &  3,6,24  &  36  &  0.1 min/155.0 Mb  &  7.2 min  &  0.5 min/126.4 Mb & 708.5 \\
5  &  1  &  3,6,24  &  36  &  0.1 min/160.3 Mb  &  7.8 min  &  0.9 min/129.8 Mb & 731.3 \\
\hline
1  &  1  &  4,8,24  &  40  &  0.1 min/171.9 Mb  &  8.5 min  &  0.7 min/136.2 Mb & 785.5 \\
2  &  1  &  4,8,24  &  40  &  0.2 min/343.1 Mb  &  17.2 min  &  1.2 min/271.9 Mb & 1567.7 \\
3  &  1  &  4,8,24  &  40  &  0.3 min/372.1 Mb  &  21.0 min  &  1.7 min/297.3 Mb & 1705.1 \\
3'  &  1  &  4,8,24  &  40  &  0.1 min/186.5 Mb  &  10.2 min  &  0.9 min/148.9 Mb & 854.6 \\
4  &  1  &  4,8,24  &  40  &  0.1 min/184.6 Mb  &  10.3 min  &  0.8 min/148.3 Mb & 846.6 \\
5  &  1  &  4,8,24  &  40  &  0.1 min/186.9 Mb  &  10.4 min  &  1.1 min/148.9 Mb & 854.5 \\
\hline
1  &  1  &  3,15,90  &  105  &  2.2 min/819.6 Mb  &  203.8 min  &  20.0 min/674.7 Mb & 3826.9 \\
2  &  1  &  3,15,90  &  105  &  4.4 min/1637.3 Mb  &  389.2 min  &  31.3 min/1352.5 Mb & 7645.2 \\
3  &  1  &  3,15,90  &  105  &  5.6 min/1945.7 Mb  &  561.2 min  &  61.1 min/821.2 Mb & 9103.8 \\
3'  &  1  &  3,15,90  &  105  &  2.9 min/974.0 Mb  &  286.7 min  &  61.1 min/410.9 Mb & 4557.2 \\
4  &  1  &  3,15,90  &  105  &  2.3 min/864.5 Mb  &  224.8 min  &  14.4 min/381.0 Mb & 4042.8 \\
5  &  1  &  3,15,90  &  105  &  3.2 min/981.1 Mb  &  291.4 min  &  342.5 min/463.4 Mb & 4571.0 \\
\hline
\hline
6  &  2  &  3,6,18  &  30  &  0.2 min/241.6 Mb  &  16.7 min  &  1.6 min/192.4 Mb & 1098.9 \\
7  &  2  &  3,6,18  &  30  &  0.2 min/244.9 Mb  &  17.3 min  &  1.8 min/194.4 Mb & 1114.4 \\
\hline
6  &  2  &  3,6,24  &  36  &  0.2 min/313.7 Mb  &  28.7 min  &  2.4 min/254.5 Mb & 1432.0 \\
7  &  2  &  3,6,24  &  36  &  0.2 min/317.6 Mb  &  31.0 min  &  2.7 min/257.5 Mb & 1450.5 \\
\hline
6  &  2  &  4,8,24  &  40  &  0.3 min/366.3 Mb  &  39.5 min  &  3.5 min/294.1 Mb & 1675.3 \\
7  &  2  &  4,8,24  &  40  &  0.3 min/371.5 Mb  &  38.2 min  &  3.8 min/297.0 Mb & 1700.1 \\
\hline
\hline
6  &  4  &  3,6,18  &  30  &  0.3 min/472.3 Mb  &  61.4 min  &  5.0 min/377.3 Mb & 2145.6 \\
7  &  4  &  3,6,18  &  30  &  0.3 min/475.5 Mb  &  62.3 min  &  5.3 min/379.3 Mb & 2161.1 \\
\hline
6  &  4  &  3,6,24  &  36  &  0.5 min/609.3 Mb  &  101.6 min  &  8.7 min/483.6 Mb & 2777.7 \\
7  &  4  &  3,6,24  &  36  &  0.5 min/613.2 Mb  &  103.1 min  &  9.2 min/486.2 Mb & 2796.2 \\
\hline
6  &  4  &  4,8,24  &  40  &  0.5 min/712.3 Mb  &  139.2 min  &  12.1 min/577.0 Mb & 3254.6 \\
7  &  4  &  4,8,24  &  40  &  0.6 min/717.5 Mb  &  141.0 min  &  12.6 min/580.3 Mb & 3279.5 \\
\hline
\end{tabular}
\end{scriptsize}
\caption{Checking properties of the communication protocol.}
\label{tab:single}
\end{center}
\end{table}

The experiments clearly shows that the formula building time is usually negligible; the satisfiability checking time is also usually acceptably small, at least within the parameter range for the experiments we considered.
On the contrary, the time to convert formulas in conjunctive normal form usually dominates in our tests.
This indicates that there is significant room for practical scalability of our verification technique.
In fact, from a computational complexity standpoint, the SAT phase is clearly the critical one, as it involves solving an NP-complete problem.
On the other hand, the CNF routine has a quadratic running time.

Another straightforward optimization could be the implementation of the TA encoding directly in CNF, to bypass the \verb+sat2cnf+ routine.
This can easily be done, because the structure of the formulas in the axiomatization is fixed.
In conclusion, we can claim safely that the performances obtained in the tests are satisfactory in perspective, and they successfully demonstrate the practical feasibility of our verification technique.

\section{Conclusion} \label{sec:conclusions}
In this paper, we introduced a verification technique to perform a partial verification of real-time systems modeled under a dense-time model and using mixed operational and descriptive components.
The technique relies on discretization techniques introduced in previous work \cite{FPR08-FM08}.
It is fully automated and implemented on top of a discrete-time bounded satisfiability checker.
We experimented with a significant example based on the description of a
communication protocol, where concurrent runs of the protocol are synchronized by means of additional MTL formulas, hence building a mixed model.
Verification tests showed consistent results and significantly good performances.


\begin{thebibliography}{10}

\bibitem{AD94}
Rajeev Alur and David~L. Dill.
\newblock A theory of timed automata.
\newblock {\em Theoretical Computer Science}, 126(2):183--235, 1994.

\bibitem{AFH96}
Rajeev Alur, Tom{\' a}s Feder, and Thomas~A. Henzinger.
\newblock The benefits of relaxing punctuality.
\newblock {\em Journal of the ACM}, 43(1):116--146, 1996.

\bibitem{AH92b}
Rajeev Alur and Thomas~A. Henzinger.
\newblock Logics and models of real time: A survey.
\newblock In J.~W. de~Bakker, Cornelis Huizing, and Willem~P. de~Roever,
  editors, {\em Proceedings of the Real-Time: Theory in Practice, REX
  Workshop}, volume 600 of {\em Lecture Notes in Computer Science}, pages
  74--106. Springer-Verlag, 1992.

\bibitem{AH93}
Rajeev Alur and Thomas~A. Henzinger.
\newblock Real-time logics: Complexity and expressiveness.
\newblock {\em Information and Computation}, 104(1):35--77, 1993.

\bibitem{BLN03}
Dirk Beyer, Claus Lewerentz, and Andreas Noack.
\newblock Rabbit: A tool for {BDD}-based verification of real-time systems.
\newblock In Warren A.~Hunt Jr. and Fabio Somenzi, editors, {\em Proceedings of
  the 15th International Conference on Computer Aided Verification (CAV'03)},
  volume 2725 of {\em Lecture Notes in Computer Science}, pages 122--125.
  Springer-Verlag, 2003.

\bibitem{Bos99}
Dragan Bo{\v s}na{\v c}ki.
\newblock Digitization of timed automata.
\newblock In {\em Proceedings of the 4th International Workshop on Formal
  Methods for Industrial Critical Systems (FMICS'99)}, pages 283--302, 1999.

\bibitem{BER94}
Ahmed Bouajjani, Rachid Echahed, and Riadh Robbana.
\newblock Verifying invariance properties of timed systems with duration
  variables.
\newblock In {\em Proceedings of the 3rd International Symposium on Formal
  Techniques in Real-Time and Fault-Tolerant Systems (FTRTFT'94)}, volume 863
  of {\em Lecture Notes in Computer Science}, pages 193--210. Springer-Verlag,
  1994.

\bibitem{BMT99}
Marius Bozga, Oded Maler, and Stavros Tripakis.
\newblock Efficient verification of timed automata using dense and discrete
  time semantics.
\newblock In Laurence Pierre and Thomas Kropf, editors, {\em Proceedings of the
  10th Correct Hardware Design and Verification Methods Advanced Research
  Working Conference (CHARME'99)}, volume 1703 of {\em Lecture Notes in
  Computer Science}, pages 125--141. Springer-Verlag, 1999.

\bibitem{CP03}
Gaurav Chakravorty and Paritosh~K. Pandya.
\newblock Digiziting interval duration logic.
\newblock In Warren~A. {Hunt, Jr.} and Fabio Somenzi, editors, {\em Proceedings
  of the 15th International Conference on Computer Aided Verification
  (CAV'03)}, volume 2725 of {\em Lecture Notes in Computer Science}, pages
  167--179. Springer-Verlag, 2003.

\bibitem{CLT07}
Edmund~M. Clarke, Flavio Lerda, and Muralidhar Talupur.
\newblock An abstraction technique for real-time verification.
\newblock In {\em Proceedings of the GM R\&D Workshop on Next Generation Design
  and Verification Methodologies for Distributed Embedded Control System},
  2007.

\bibitem{DWDR05}
Martin {De Wulf}, Laurent Doyen, and Jean-Fran{\c c}ois Raskin.
\newblock Almost {ASAP} semantics: from timed models to timed implementations.
\newblock {\em Formal Aspects of Computing}, 17(3):319--341, 2005.

\bibitem{DMP06}
D.~D'Souza, R.~{Mohan M.}, and P.~Prabhakar.
\newblock Eliminating past operators in metric temporal logic.
\newblock Technical Report IISc-CSA-TR-2006-11, 2006.

\bibitem{FP07}
Georgios~E. Fainekos and George~J. Pappas.
\newblock Robust sampling for {MITL} specifications.
\newblock In {\em Proc. of FORMATS'07}, volume 4763 of {\em LNCS}, 2007.

\bibitem{Fre05}
Goran Frehse.
\newblock {PHAVer}: Algorithmic verification of hybrid systems past {HyTech}.
\newblock In {\em Proceedings of the 5th International Workshop on Hybrid
  Systems: Computation and Control (HSCC'05)}, volume 3414 of {\em Lecture
  Notes in Computer Science}, pages 258--273. Springer-Verlag, 2005.

\bibitem{FMMR07-TR2007-22}
Carlo~A. Furia, Dino Mandrioli, Angelo Morzenti, and Matteo Rossi.
\newblock Modeling time in computing: a taxonomy and a comparative survey.
\newblock Technical Report 2007.22, Dipartimento di Elettronica e Informazione,
  Politecnico di Milano, January 2007.

\bibitem{FPR08-FM08}
Carlo~A. Furia, Matteo Pradella, and Matteo Rossi.
\newblock Automated verification of dense-time {MTL} specifications via
  discrete-time approximation.
\newblock In Jorge Cu{\'e}llar and Tom Maibaum, editors, {\em Proceedings of
  the 15th International Symposium on Formal Methods (FM'08)}, volume 5014 of
  {\em Lecture Notes in Computer Science}, pages 132--147. Springer-Verlag, May
  2008.

\bibitem{FR06}
Carlo~A. Furia and Matteo Rossi.
\newblock Integrating discrete- and continuous-time metric temporal logics
  through sampling.
\newblock In Eugene Asarin and Patricia Bouyer, editors, {\em Proceedings of
  the 4th International Conference on Formal Modelling and Analysis of Timed
  Systems (FORMATS'06)}, volume 4202 of {\em Lecture Notes in Computer
  Science}, pages 215--229. Springer-Verlag, September 2006.

\bibitem{FR07-FORMATS07}
Carlo~A. Furia and Matteo Rossi.
\newblock On the expressiveness of {MTL} variants over dense time.
\newblock In Jean-Fran{\c c}ois Raskin and P.~S. Thiagarajan, editors, {\em
  Proceedings of the 5th International Conference on Formal Modelling and
  Analysis of Timed Systems (FORMATS'07)}, volume 4763 of {\em Lecture Notes in
  Computer Science}, pages 163--178. Springer-Verlag, October 2007.

\bibitem{Fur07}
Carlo~Alberto Furia.
\newblock {\em Scaling up the formal analysis of real-time systems}.
\newblock PhD thesis, Dipartimento di Elettronica e Informazione, Politecnico
  di Milano, May 2007.

\bibitem{GPV94}
Aleks G{\" o}ll{\" u}, Anuj Puri, and Pravin Varaiya.
\newblock Discretization of timed automata.
\newblock In {\em Proceedings of the 33rd Conference on Decision and Control},
  pages 957--958, 1994.

\bibitem{HHW97}
Thomas~A. Henzinger, Pei-Hsin Ho, and Howard Wong-Toi.
\newblock {HYTECH}: A model checker for hybrid systems.
\newblock {\em International Journal on Software Tools for Technology
  Transfer}, 1(1--2), 1997.

\bibitem{HMP92}
Thomas~A. Henzinger, Zohar Manna, and Amir Pnueli.
\newblock What good are digital clocks?
\newblock In Werner Kuich, editor, {\em Proceedings of the 19th International
  Colloquium on Automata, Languages and Programming (ICALP'92)}, volume 623 of
  {\em Lecture Notes in Computer Science}, pages 545--558. Springer-Verlag,
  1992.

\bibitem{HRS98}
Thomas~A. Henzinger, Jean-Fran{\c c}ois Raskin, and Pierre-Yves Schobbens.
\newblock The regular real-time languages.
\newblock In Kim~Guldstrand Larsen, Sven Skyum, and Glynn Winskel, editors,
  {\em Proceedings of the 25th International Colloquium on Automata, Languages
  and Programming (ICALP'98)}, volume 1443 of {\em Lecture Notes in Computer
  Science}, pages 580--591. Springer-Verlag, 1998.

\bibitem{VHG96}
Dang~Van Hung and Phan~Hong Giang.
\newblock Sampling semantics of {D}uration {C}alculus.
\newblock In Joachim~Parrow Bengt~Jonsson, editor, {\em Proceedings of the 4th
  International Symposium on Formal Techniques in Real-Time and Fault-Tolerant
  Systems (FTRTFT'96)}, volume 1135 of {\em Lecture Notes in Computer Science},
  pages 188--207. Springer-Verlag, 1996.

\bibitem{Koy90}
Ron Koymans.
\newblock Specifying real-time properties with metric temporal logic.
\newblock {\em Real-Time Systems}, 2(4):255--299, 1990.

\bibitem{KP05}
Pavel Kr{\v c}{\' a}l and Radek Pel{\' a}nek.
\newblock On sampled semantics of timed systems.
\newblock In R.~Ramanujam and Sandeep Sen, editors, {\em Proceedings of the
  25th International Conference on Foundations of Software Technology and
  Theoretical Computer Science (FSTTCS'05)}, volume 3821 of {\em Lecture Notes
  in Computer Science}, pages 310--321. Springer-Verlag, 2005.

\bibitem{LPY97}
Kim~G. Larsen, Paul Pettersson, and Wang Yi.
\newblock {UPPAAL} in a nutshell.
\newblock {\em International Journal on Software Tools for Technology
  Transfer}, 1(1--2), 1997.

\bibitem{MNP06}
Oded Maler, Dejan Nickovic, and Amir Pnueli.
\newblock From {MITL} to timed automata.
\newblock In Eugene Asarin and Patricia Bouyer, editors, {\em Proceedings of
  the 4th International Conference on Formal Modeling and Analysis of Timed
  Systems (FORMATS'06)}, volume 4202 of {\em Lecture Notes in Computer
  Science}, pages 274--289. Springer-Verlag, 2006.

\bibitem{MP95}
Oded Maler and Amir Pnueli.
\newblock Timing analysis of asynchronous circuits using timed automata.
\newblock In Paolo Camurati and Hans Eveking, editors, {\em Proceedings of the
  Advanced Research Working Conference on Correct Hardware Design and
  Verification Methods}, volume 987 of {\em Lecture Notes in Computer Science},
  pages 189--205. Springer-Verlag, 1995.

\bibitem{Oua02}
Jo{\" e}l Ouaknine.
\newblock Digitisation and full abstraction for dense-time model checking.
\newblock In Joost-Pieter Katoen and Perdita Stevens, editors, {\em Proceedings
  of the 8th International Conference on Tools and Algorithms for the
  Construction and Analysis of Systems (TACAS'02)}, volume 2280 of {\em Lecture
  Notes in Computer Science}, pages 37--51. Springer-Verlag, 2002.

\bibitem{OW03}
Jo{\" e}l Ouaknine and James Worrell.
\newblock Revisiting digitization, robustness, and decidability for timed
  automata.
\newblock In {\em Proceedings of the 18th Annual IEEE Symposium on Logic in
  Computer Science (LICS'03)}, pages 198--207. IEEE Computer Society Press,
  2003.

\bibitem{zot}
Matteo Pradella.
\newblock \zot{}.
\newblock \texttt{http://home.dei.polimi.it/pradella}, March 2007.

\bibitem{PMS07}
Matteo Pradella, Angelo Morzenti, and Pierluigi {San Pietro}.
\newblock The symmetry of the past and of the future: bi-infinite time in the
  verification of temporal properties.
\newblock In {\em Proc. of ESEC/FSE 2007}, 2007.

\bibitem{SPC05}
Babita Sharma, Paritosh~K. Pandya, and Supratik Chakraborty.
\newblock Bounded validity checking of interval duration logic.
\newblock In Nicolas Halbwachs and Lenore~D. Zuck, editors, {\em Proceedings of
  the 11th International Conference on Tools and Algorithms for the
  Construction and Analysis of Systems (TACAS'05)}, volume 3440 of {\em Lecture
  Notes in Computer Science}, pages 301--316. Springer-Verlag, 2005.

\bibitem{Yov97}
Sergio Yovine.
\newblock Kronos: A verification tool for real-time systems.
\newblock {\em International Journal on Software Tools for Technology
  Transfer}, 1(1--2):123--133, 1997.

\end{thebibliography}

\end{document}